\let\includefigures=\iftrue
%
\let\useblackboard=\iftrue
%
%
\newfam\black
\input harvmac
\noblackbox
\includefigures
\message{If you do not have epsf.tex (to include figures),}
\message{change the option at the top of the tex file.}
\input epsf
\def\figin{\epsfcheck\figin}\def\figins{\epsfcheck\figins}
\def\epsfcheck{\ifx\epsfbox\UnDeFiNeD
\message{(NO epsf.tex, FIGURES WILL BE IGNORED)}
\gdef\figin##1{\vskip2in}\gdef\figins##1{\hskip.5in}
\else\message{(FIGURES WILL BE INCLUDED)}%
\gdef\figin##1{##1}\gdef\figins##1{##1}\fi}
\def\DefWarn#1{}
\def\figinsert{\goodbreak\midinsert}
\def\ifig#1#2#3{\DefWarn#1\xdef#1{fig.~\the\figno}
\writedef{#1\leftbracket fig.\noexpand~\the\figno}%
\figinsert\figin{\centerline{#3}}\medskip\centerline{\vbox{
\baselineskip12pt\advance\hsize by -1truein
\noindent\footnotefont{\bf Fig.~\the\figno:} #2}}
\bigskip\endinsert\global\advance\figno by1}
\else
\def\ifig#1#2#3{\xdef#1{fig.~\the\figno}
\writedef{#1\leftbracket fig.\noexpand~\the\figno}%
\global\advance\figno by1}
\fi
%

\def\smallfig#1#2#3{\DefWarn#1\xdef#1{fig.~\the\figno}
\writedef{#1\leftbracket fig.\noexpand~\the\figno}%
\figinsert\
in{\centerline{#3}}\medskip\centerline{\vbox{
\baselineskip12pt\advance\hsize by -1truein
\noindent\footnotefont{\bf Fig.~\the\figno:} #2}}
\endinsert\global\advance\figno by1}

\useblackboard
\message{If you do not have msbm (blackboard bold) fonts,}
\message{change the option at the top of the tex file.}
\font\blackboard=msbm10 scaled \magstep1
\font\blackboards=msbm7
\font\blackboardss=msbm5
\textfont\black=\blackboard
\scriptfont\black=\blackboards
\scriptscriptfont\black=\blackboardss

\else

\fi
%


\def\boxit#1{\vbox{\hrule\hbox{\vrule\kern8pt
\vbox{\hbox{\kern8pt}\hbox{\vbox{#1}}\hbox{\kern8pt}}
\kern8pt\vrule}\hrule}}
\def\mathboxit#1{\vbox{\hrule\hbox{\vrule\kern8pt\vbox{\kern8pt
\hbox{$\displaystyle #1$}\kern8pt}\kern8pt\vrule}\hrule}}

\def\subsubsection#1{\bigskip\noindent
{\it #1}}

\def\yboxit#1#2{\vbox{\hrule height #1 \hbox{\vrule width #1
\vbox{#2}\vrule width #1 }\hrule height #1 }}
\def\fillbox#1{\hbox to #1{\vbox to #1{\vfil}\hfil}}
\def\ybox{{\lower 1.3pt \yboxit{0.4pt}{\fillbox{8pt}}\hskip-0.2pt}}
%
%



\def\l{\left}

\def\comments#1{}

\def\p{\partial}

\def\half{{1\over 2}}

\def\CM{{\cal M}}
\def\CN{{\cal N}}
\def\CO{{\cal O}}

\def\CL{{\cal L}}

\def\CX{{\cal X}}

\def\ap{\alpha'}

\def\II{\relax{I\kern-.10em I}}

\font\cmss=cmss10 \font\cmsss=cmss10 at 7pt
\def\IZ{\relax\ifmmode\mathchoice
{\hbox{\cmss Z\kern-.4em Z}}{\hbox{\cmss Z\kern-.4em Z}}
{\lower.9pt\hbox{\cmsss Z\kern-.4em Z}}
{\lower1.2pt\hbox{\cmsss Z\kern-.4em Z}}
\else{\cmss Z\kern-.4emZ}\fi}
\def\IR{\relax{\rm I\kern-.18em R}}
\def\IZ{\relax\ifmmode\mathchoice
{\hbox{\cmss Z\kern-.4em Z}}{\hbox{\cmss Z\kern-.4em Z}}
{\lower.9pt\hbox{\cmsss Z\kern-.4em Z}} {\lower1.2pt\hbox{\cmsss
Z\kern-.4em Z}}\else{\cmss Z\kern-.4em Z}\fi}
\def\IB{\relax{\rm I\kern-.18em B}}
\def\IC{{\relax\hbox{$\inbar\kern-.3em{\rm C}$}}}
\def\ID{\relax{\rm I\kern-.18em D}}
\def\IE{\relax{\rm I\kern-.18em E}}
\def\IF{\relax{\rm I\kern-.18em F}}
\def\IG{\relax\hbox{$\inbar\kern-.3em{\rm G}$}}
\def\IGa{\relax\hbox{${\rm I}\kern-.18em\Gamma$}}
\def\IH{\relax{\rm I\kern-.18em H}}
\def\II{\relax{\rm I\kern-.18em I}}
\def\IK{\relax{\rm I\kern-.18em K}}
\def\IP{\relax{\rm I\kern-.18em P}}

%

\def\inbar{\,\vrule height1.5ex width.4pt depth0pt}

\def\p{\partial}

\font\cmss=cmss10 
\def\IR{\relax{\rm I\kern-.18em R}}

%


%

\def\ms{m_s}
\def\gs{g_s}
\def\lp10{\ell_p^{10}}
\def\lp11{\ell_p^{11}}
\def\R11{R_{11}}

\def\frac#1#2{{#1 \over #2}}


\def\l{\left}

\def\comments#1{}

\def\p{\partial}

\def\half{{1\over 2}}

\def\CM{{\cal M}}
\def\CN{{\cal N}}
\def\CO{{\cal O}}

\def\CL{{\cal L}}

\def\CX{{\cal X}}

\def\Hor{Ho\v{r}ava}
\def\ibid{{\it ibid.}}
\def\cf{{\it c.f.}}
\def\M4{M_{Pl,4}}
\def\rpm{m_{Pl,4}}
\def\mg{M_{GUT}}
\def\ag{\alpha_{GUT}}
\def\k11{\kappa_{11}}
\def\l11{\ell_{11}}
\def\tl11{\tilde{\ell}_{11}}
\def\sqg{\sqrt{g}}
\def\m11{M_{11}}
\def\tm11{\tilde{M}_{11}}

\def\np{{\it Nucl. Phys.}}
\def\prl{{\it Phys. Rev. Lett.}}
\def\pr{{\it Phys. Rev.}}
\def\pl{{\it Phys. Lett.}}

\lref\kolbturner{E.W. Kolb and M.S. Turner, {\it The Early Universe}
Addison-Wesley (1990).}
\lref\lindebook{A. Linde,
{\it Particle Physics and Inflationary Cosmology}
Harwood Academic, (1990).}

\lref\liddlebook{A. Liddle and D. Lyth, {\it Cosmological Inflation
and Large-Scale Structure} Cambridge University Press (2000).}

\lref\reconsrev{For a review see
J. E. Lidsey {\it et. al.},
``Reconstructing the Inflaton Potential--An Overview,"
Reviews of Modern Physics, {\bf 69} (1997) 373.}
\lref\lindehybrid{A. Linde, ``Axions in Inflationary Cosmology,"
\pl {\bf B259} (1991) 38; {\it ibid.},
``Hybrid Inflation," \pr {\bf D49} (1994) 748, astro-ph/9307002.}

\lref\naturalinfl{K. Freese, J.A. Frieman and A.V. Olinto,
``Natural Inflation with Pseudo-Nambu-Goldstone Bosons",
Phys. Rev. Lett. {\bf 65} (1990) 3233;
F.C. Adams, J.R. Bond, K. Freese, J.A. Frieman and A.V. Olinto,
``Natural Inflation:  Particle Physics Models,
Power Law Spectra for Large Scale Structure, and Constraints from
COBE", Phys. Rev. {\bf D47} (1993) 426, hep-ph/9207245}

\lref\kolbresonant{D.J.H. Chung,
E.W. Kolb, A. Riotto and I. Tkachev,
``Probing Planckian Physics: Resonant
Production of Particles During Inflation and
Features in the Primordial Power Spectrum,"
Phys.Rev. {\bf D62} (2000) 043508,
hep-ph/9910437.}

\lref\martinbrand{J. Martin and R. H. Brandenberger,
``The Transplanckian Problem of Inflationary
 Cosmology," Phys. Rev. {\bf D63}(2001) 123501, hep-th/0005209; R. H.
Branbenberger and J. Martin,
 ``The Robustness of Inflation to Changes
in Superplanck Scale Physics," Mod. Phys. Lett. {\bf A16} (2001)
999,  astro-ph/0005432; R. H. Brandenberger, S. E. Joras, and J. Martin,
Trans-Planckian Physics and the Spectrum
of Fluctuations in a Bouncing Universe, hep-th/0112122 . }

\lref\Niem{J. C. Niemeyer, ``Inflation With a Planck Scale
Frequency Cutoff," Phys. Rev. {\bf D63} (2001) 123502,
astro-ph/0005533; J. C. Niemeyer and R. Parentani,
``Transplanckian Dispersion and Scale invariance
of Inflationary Perturbations,"
Phys. Rev. {\bf D64} (2001) 101301,  astro-ph/0101451.}

\lref\kempf{A. Kempf and J. C. Niemeyer,
``Perturbation Spectrum in Inflation With Cutoff,"
Phys. Rev. {\bf D64} (2001) 103501,
 astro-ph/0103225.}

\lref\egks{R. Easther, B. R. Greene, W. H. Kinney,
and G. Shiu, ``Inflation as a Probe
of Short Distance Physics, " Phys. Rev. {\bf D64} (2001) 103502, hep-th/01044102.}

\lref\egkss{R. Easther, B. R. Greene,
W. H. Kinney, and G. Shiu, ``Imprints of Short Distance
Physics on Inflationary Cosmology," hep-th/0110226 .}

\lref\witstrong{E. Witten, ``Strong Coupling
Expansion of Calabi-Yau Compactification," \np\ {\bf B471} (1996) 135,
 hep-th/9602070}

\lref\conscon{L. Hui and W. H. Kinney,
``Short Distance Physics and the Consistency Relation
for Scalar and Tensor Fluctuations in the Inflationary Universe,"
astro-ph/0109107.  For examples of other models that discuss
modifications of the consistency conditions, see {\it e.g.},
J. Garriga and V. F. Mukhanov, ``Perturbations in K-Inflation,"
Phys. Lett. {\bf B458} (1999) 219, hep-th/9904176; and
G. Shiu and S. H. H. Tye, ``Some Aspects of Brane Inflation,"
Phys. Lett. {\bf B516} (2001) 421, hep-th/0106274.}

\lref\wang{L. Wang, V. Mukhanov, and P. Steinhardt, ``On the Problem of
Predicting Inflationary Perturbations," Phys. Lett. {\bf B414} (1997) 18,  astro-ph/9709032.}

\lref\bst{J. Bardeen, P. Steinhardt, and M. Turner, ``Spontaneous Creation of Almost Scale-Free
Density Perturbations in an Inflationary Universe," Phys. Rev {\bf D28} (1983) 679.}

\lref\cgkl{E. Copeland, I. Girvell, E. Kolb, and A. Liddle, ``On the Reliability of Inflaton
Potential Reconstruction," Phys. Rev. {\bf D58} (1998) 043002, astro-ph/9802209.}

\lref\cobe{J. C. Mather {\it et al.}, ``A Preliminary Measurement of the Cosmic Microwave 
Background Spectrum by the Cosmic Microwave Background Explorer (COBE) Satellite,"
Astrophys. J. {\bf 354} (1990) L37.} 

\lref\boomerang{A. E. Lange {\it et al.}, ``Cosmological Parameters from the First Results
of BOOMERANG," Phys. Rev. {\bf D63} (2001) 042001, astro-ph/0005004.}

\lref\maxima{Balbi, A. {\it et al.}, ``Constraints on Cosmological Parameters from MAXIMA-1",
 Astrophys. J. {\bf 545} (2000) L5, astro-ph/0005124.}
    

\lref\bankscosmo{T. Banks, ``Cosmological Breaking of Supersymmetry?
or Little Lambda Goes Back to the Future 2. ," hep-th/0007146 .}

\lref\selzal{U. Seljak and M. Zaldarriaga,
``Signature of Gravity Waves in Polarization of the Microwave Background,"
{\it Phys.Rev.Lett.} {\bf 78} (1997) 2054.}

\lref\fs{W. Fischler and L. Susskind, ``Holography and Cosmology,''
hep-th/9806039.}
\lref\rb{R. Bousso, ``A Covariant Entropy Conjecture,''
JHEP {\bf 9907} (1999) 004,
hep-th/9905177; {\it ibid.},
``Holography in general space-times,''
JHEP {\bf 9906} (1999) 028,
hep-th/9906022.}
\lref\hks{S. Hellerman, N. Kaloper and L. Susskind,
``String theory and quintessence,''
JHEP {\bf 0106}, (2001) 003,
hep-th/0104180;
W. Fischler, A. Kashani-Poor, R. McNees and S. Paban,
``The Acceleration of the Universe, a Challenge for String Theory",
JHEP {\bf 0107} (2001) 003, hep-th/0104181.}
\lref\Bard{J. Bardeen,
``Gauge Invariant Cosmological Perturbations,''
\pr\ {\bf D22}\ (1980) 1882.}
\lref\horwit{P. \Hor\ and E. Witten,
``Heterotic and Type I string dynamics
form eleven dimensions'', \np\ {\bf B460}\
(1996) 506, hep-th/9510209; \ibid, ``Eleven-dimensional
supergravity on a manifold with boundary'',
\np\ {\bf B475} (1996) 94, hep-th/9603142.}
\lref\add{N. Arkani-Hamed, S. Dimopoulos
and G.R. Dvali, ``The Hierarchy Problem and
New Dimensions at a Millimeter,''
\pl\ {\bf B429} (1998) 263,
hep-ph/9803315; \ibid,
``Phenomenology, Astrophysics and Cosmology
of Theories with Sub-millimeter Dimensions and TeV
Scale Quantum Gravity,''
\pr\ {\bf D59}\ (1999) 086004,
hep-ph/9807344.}
\lref\kalin{N. Kaloper and A. Linde,
``Inflation and Large Internal Dimensions,''
\pr\ {\bf D59} (1999) 101303,
hep-th/9811141.}
\lref\addcos{N. Arkani-Hamed, S. Dimopoulos,
N. Kaloper and J. March-Russell,
``Rapid Asymmetric Inflation and
Early Cosmology in Theories with Sub-millimeter Dimensions,''
\np\ {\bf B567} (2000) 189,
hep-ph/9903224.}

\lref\unify{S. Dimopoulos and H. Georgi, ``Softly Broken Supersymmetry
and SU(5)", Nucl. Phys. {\bf B192}
(1981) 150; S. Dimopoulos, S. Raby and F. Wilczek, ``Supersymmetry and
the Scale of Unification", Phys. Rev. {\bf
D24} (1981) 1681;
U. Amaldi, W. de Boer and H. Furstenau,
``Comparison of Grand Unified Theories with Electroweak and Strong Coupling
Constants Measured at LEP",
Phys. Lett. {\bf B260} (1991) 447;
P. Langacker and N. Polonsky,
``Uncertainties in Coupling Constant Unification",
Phys. Rev. {\bf D47} (1993) 4028, hep-ph/9210235.}
\lref\linde{A.D. Linde, ``Chaotic Inflation,''
\pl\ {\bf B129} (1983) 177.}
\lref\gsw{M.B. Green, J. Schwarz and E. Witten,
{\it Superstring Theory}, vols. I and II,
Cambridge Univ. Press (1987).}
\lref\gluino{P. \Hor,
``Gluino Condensation in Strongly Coupled
Heterotic String Theory,''
\pr\ {\bf D54} (1996) 7561,
hep-th/9608019.}
\lref\tomcosm{T. Banks, ``Remarks on M Theoretic
Cosmology,'' hep-th/9906126.}
\lref\tomcosmrev{T. Banks, ``M-theory and cosmology,'' hep-th/9911067.}
\lref\feynman{R.P. Feynman et. al., {\it Feynman
Lectures on Gravitation}}

\lref\kamkos{M. Kamionkowski and A. Kosowsky,
``The Cosmic Microwave Background and Particle Physics,''
{\it Ann.\ Rev.\ Nucl.\ Part.\ Sci.}  {\bf 49} (1999) 77,
astro-ph/9904108.}
\lref\gtwofiber{J.~A.~Harvey, D.~A.~Lowe and A.~Strominger,
``N=1 String Duality,''
\pl\ {\bf B362} (1995) 65,
hep-th/9507168; B.~S.~Acharya,
``N=1 Heterotic/M-theory Duality and Joyce Manifolds,''
\np\ {\bf B475} (1996) 579,
hep-th/9603033.}
\lref\gtwochiral{E.~Witten,
``Anomaly Cancellation on $G_2$ Manifolds,''
hep-th/0108165; B.~S.~Acharya and E.~Witten,
``Chiral Fermions from Manifolds of $G_2$ Holonomy,''
hep-th/0109152; E. Witten, ``Deconstruction,
$G_2$ Holonomy, and Doublet-Triplet
Splitting,'' hep-ph/0201018.}

\lref\syz{A. Strominger, S.-T. Yau and E. Zaslow,
``Mirror Symmetry is T-duality,''
\np\ {\bf B479}\ (1996) 243, hep-th/9606040.}
\lref\addcosmprob{K. Benakli and S. Davidson,
``Baryogenesis in Models with a Low Quantum Gravity Scale,''
\pr\ {\bf D60} (1999) 025004, hep-ph/9810280;
D. Lyth, ``Inflation With TeV-scale Gravity Needs
Supersymmetry,'' \pl\ {\bf B448}\ (1999) 191,
hep-ph/9810320.}

\lref\modular{
P. Binetruy and M.K. Gaillard,
``Candidates for the Inflaton Field in Superstring Models,"
Phys. Rev. D34 (1986) 3069;
 L. Randall and S. Thomas ,
``Solving the Cosmological Moduli Problem with Weak Scale Inflation,"
 Nucl. Phys. {\bf B449} (1995) 229,
hep-ph/9407248;
T. Banks, M. Berkooz, G. Moore, S. Shenker
and P. Steinhardt, ``Modular cosmology,''
\pr\ {\bf D52} (1995) 3548, hep-th/9503114;
S. Thomas,
``Moduli Inflation from Dynamical Supersymmetry Breaking,"
Phys.Lett. {\bf B351} (1995) 424,
hep-th/9503113.}

\lref\ruts{M. Green and P. Vanhove, ``D Instantons, Strings, and M-Theory,"
Phys. Lett. {\bf B 408} (1997) 122, hep-th/9704145;
M. B. Green, M. Gutperle, and P. Vanhove, ``One Loop in Eleven--Dimensions,"
Phys. Lett. {\bf B409} (1997) 177, hep-th/9706175; and J. Russo and A. Tseytlin,
``One-loop Four-graviton Amplitude in
Eleven-Dimensional Supergravity,''
Nuc. Phys. {\bf B508} (1997) 245, hep-th/9707134.}

\lref\hettypeI{J. Polchinski and E. Witten,
``Evidence for Heterotic - Type I String Duality,''
\np\ {\bf B460} (1996) 525, hep-th/9510169.}

\lref\dsscales{M. Dine and N. Seiberg, ``Couplings and
Scales in Superstring Models,'' \prl\ {\bf 55}\
(1985) 366.}
\lref\kscales{V. Kaplunovsky, ``Mass
Scales of the String Unification,''
\prl\ {\bf 55}\ (1985) 1036.}
\lref\multif{D. Polarski and A.A. Starobinsky,
``Structure of Primordial Gravitational
Waves Spectrum in a Double Inflationary
Model,'' \pl\ {\bf B356}\ (1995) 196,
astro-ph/9505125; J. Garcia-Bellido and
D. Wands, \pl\ {\bf D52}\ (1995) 6739;
M. Sasaki and E. Stewart, ``A General Analytic
Formula for the Spectral Index of the Density
Perturbations Produced During Inflation,''
{\it Prog. Theor. Phys.}\ {\bf 95}\ (1996) 71,
astro-ph/9507001.}
\lref\corrconsist{N. Bartolo, S. Matarrese and
A. Riotto, ``Adiabatic and Isocurvature Perturbations
from Inflation: Power Spectra and Consistency
Relations,'' \pr\ {\bf D64} (2001) 123504.}
\lref\isoc{A.D. Linde, ``Generation of
Isothermal Density Perturbations in the
Inflationary Universe,'' \pl\ {\bf B158} (1985) 375;
L.A. Kofman, ``What Initial Perturbations may be
Generated in Inflationary Cosmological Models,''
\pl\ {\bf B173}\ (1986) 400; L.A. Kofman and
A. Linde, ``Generation of Density Perturbations
in Inflationary Cosmology,'' \np\ {\bf B282}\ (1987) 555;
A. Linde and V. Mukhanov, ``Non-Gaussian Isocurvature
Perturbations from Inflation,'' \pr {\bf D56} (1997) R535;
P.J.E. Peebles, astro-ph/9805194.}
\lref\efbond{G. Efstathiou and J.R. Bond,
{\it Mon. Not. R. Astron. Soc.} {\bf 218}\
(1986) 103.}
\lref\langlois{D. Langlois, ``Correlated Adiabatic and
Isocurvature Perturbations from Double Inflation,''
\pr\ {\bf D59} (1999) 123512; D. Langlois and
A. Riazuelo, ``Correlated Mixtures of Adiabatic and
Isocurvature Cosmological Perturbations,''
\pr\ {\bf D62}\ (2000) 043504; C. Gordon {\it et. al.},
``Adiabatic and Entropy Perturbations from Inflation,''
\pr\ {\bf D63} (2000) 023506; N. Bartolo, S. Mattarese
and A. Riotto, ``Oscillations During Inflation and the
Cosmological Density Perturbations,'' \pr\ {\bf D64}\
(2001) 083514.}
\lref\correxp{M. Bucher, K. Moodley and N. Turok,
``General Primordial Cosmic Perturbation,'' \pr\ {\bf D62}
(2000) 083508; R. Trotta, A. Riazuelo and R. Durrer,
``Cosmic Microwave Background Anisotropies with
Mixed Isocurvature Perturbations,'' astro-ph/0104107;
L. Amendola {\it et. al.}, ``Correlated Perturbations
{}From Inflation and the Cosmic Microwave Background,''
astro-ph/0107089.}
\lref\huandwhite{For a pedagogical review see
W.~Hu and M.~J.~White,
``A CMB Polarization Primer,''
{\it New Astron.}  {\bf 2} (1997) 323,
astro-ph/9706147.}


\Title{\vbox{\baselineskip12pt\hbox{hep-th/0201158}
\hbox{SU-ITP-02/02} \hbox{SLAC-PUB-9112}}} {\vbox{
\centerline{Signatures of Short Distance Physics in the}
\smallskip \centerline{Cosmic Microwave Background}
}}
\smallskip
\centerline{Nemanja Kaloper$^{1}$, Matthew Kleban$^{1}$,
Albion Lawrence$^{1,2}$ and Stephen Shenker$^{1}$}
\bigskip
\bigskip
\centerline{$^{1}${Department of Physics,
Stanford University, Stanford, CA 94305}}
\medskip
\centerline{$^{2}${SLAC Theory Group, MS 81, 2575 Sand Hill Road,
Menlo Park, CA 94025}}
\bigskip
\bigskip
\noindent

We systematically investigate the effect of short distance physics on
the spectrum of temperature anistropies in the Cosmic Microwave Background
produced during
inflation.  We present
a general argument--assuming only low energy locality--that the size of such
effects are
of order $H^2/M^2$, where $H$ is the Hubble parameter during inflation, and
$M$ is the scale
of the high energy physics.

We evaluate the strength of such effects in a number of specific string and
M theory models.
In weakly coupled field theory and string theory models,
the effects are far too small to be observed.
In phenomenologically attractive \Hor-Witten
compactifications, the effects are much larger but still
unobservable. In certain
M theory models, for which
the fundamental Planck scale is several orders of
magnitude below the conventional scale of
grand unification, the effects may be
on the threshold of detectability.

However, observations of both
the scalar and tensor fluctuation contributions to the Cosmic Microwave Background
power spectrum--with a precision near the cosmic variance limit--are necessary in order
to unambiguously demonstrate the
existence of these signatures of high energy physics.  This is a formidable
experimental challenge.

\medskip
\bigskip

\Date{January 2002}

\newsec{Introduction}
The enormous disparity in scales between the observed Planck mass
$\sim 10^{19}$ GeV and the energy
of current accelerators  ($10^{3}$ GeV) stands
as the main barrier to connecting theoretical work in quantum gravity to
experiment.   There are a few
exceptions to this difficult situation.  Proton decay experiments
overcome this immense disparity by examining decays
in kilotons of protons for millions of seconds.
Investigations of coupling constant unification
use the slow, logarithmic variation of couplings combined with
the assumption of a desert
to extract information
about the nature and scale of unification.  But such bright spots are few
and far between.

Observational cosmology provides a window into very early times and hence,
most think, into very high energy processes.  This possible
high energy probe has received much more attention recently because of the
new data available, the experiments being done, and the experiments
being planned to study the cosmic microwave backround radiation (CMBR).
The benchmark theory that explains the fluctuations
in the CMBR is  inflation\foot{ For textbook introductions see
 \lindebook \kolbturner \liddlebook.},
which traces them to ``thermal" quanta of a scalar
inflaton field during a time of exponential expansion
of the universe.  In the simplest models of inflation the scale of vacuum energy
during this period of exponential expansion was $\sim 10^{16}$ GeV
and the rate of exponential expansion $H \sim 10^{13}-10^{14}$ GeV.  These enormous
energies suggest that during the inflationary epoch
various kinds of high energy processes were activated, and further, that they
could have left their imprint on the CMBR.

Many authors have drawn attention to this exciting prospect.  The first piece of
high energy physics to
be unraveled could well be dynamics of inflation itself.  Much work
has gone into how to
reconstruct the potential of the inflaton field from CMBR data \reconsrev .

We should stress at this
point that it is by no means necessary for the scale of inflation to be as
high as $H \sim
10^{13}-10^{14}$ GeV. Other inflationary
models, e.g. hybrid models  \lindehybrid,
exist where $H$ can be
much lower, for example, $H \sim 10^3$ GeV.
Fortunately the scale of inflation can
be experimentally
tested.  Since gravity couples to mass-energy, the amount of gravitational
radiation produced
during inflation is directly related to the energy available during
inflation.  This
gravitational radiation imprints itself as a polarized component of the CMBR,
whose power is
proportional to $(H/m_4)^2$, where $m_4$ is the (reduced) four dimensional Planck length.
So measurements of this power give a direct
measurement of $H$.
COBE \cobe\ data already provide the
interesting upper bound $H < 10^{14}$ GeV which corresponds to vacuum energies
$\sim  10^{16}$ GeV, the supersymmetric unification scale.
Intensive efforts are under
way to improve this measurement.

In this paper we will concentrate on the ``high scale" possibility for $H$
since this gives the largest range for discovering new
physics via inflation.   There have been a number of investigations of the
signature of high energy scale physics in the CMBR.  Heavy particles
produced by parametric resonance have been studied in \kolbresonant .
Several groups \refs{\martinbrand,\Niem,\kempf,\egks,\egkss}
have studied the effect that
simple phenomenological models of
stringy corrections to gravity would have on the inflationary fluctuation
spectrum in the CMBR.  This work shares many
features with the results we will present.
These groups found that the size of these
imprints on the CMBR is controlled by the
natural dimensionless ratio  $r = H^2/m_s^2$
where $m_s$ is the string mass.  For conventional weakly coupled  string
theories containing gravity $m_s$ is approximately the same as the four
dimensional Planck mass $\sim 10^{19}$ GeV so $r \sim 10^{-11}$.
The actual size of the effects in these
models depends on some delicate issues of boundary
conditions at short distances that are not completely specified by the model.
These groups have surveyed the range of possible long distance behaviors allowed by different
boundary conditions. The authors of \kempf\ have focused on boundary conditions that yield
imprints of size
$\sim r$ while the authors  of \egkss\ have focused on boundary conditions yielding effects of
of size  $\sim r^{n}, n\sim .5$  .
The
analysis we present below shows that the effects are of size $\sim r$ in any theory that
is local on momentum scales $\le H$, an apparently sensible physical requirement.
Such an effect is far too small to observe for $r \sim 10^{-11}$.
In fact, the ultimate statistical limit
of cosmic variance, the number of independent sky samples available,
excludes it from being observed even in principle.

It is important to note, though,  how
great an improvement this ratio is over the suppression accelerator
based physicists must confront.  The energies accessible to them are of
order $10^3$ GeV so their suppressions are of order
$ (10^3/10^{19})^2 \sim 10^{-32}$.   But
the fact that $r \sim 10^{-11}$ is a vast improvement is cold comfort to an
experimentalist waiting for counts in an apparatus.

But,  as  pointed out in \egks ,  modern string and M theory
models  allow for the possibility of lower values of the
fundamental mass scales, raising
the possibility of more favorable ratios $r$.  Much of this paper will be
devoted to exploring this question in detail.

In Section 2, we will briefly review the framework of slow roll inflation,
explaining the basic observable quantities in both scalar and
tensor fluctuations. We emphasize that the size of inflationary perturbations is
fully determined by physics at the scale  $H $ which is much below
the Planck scale. Therefore the locality of effective theory used to compute
the fluctuations implies that these perturbations are independent of the  details
of Planck scale physics.

In Section 3, we will explain the basic mechanism by which  high
energy physics leaves an imprint on CMBR fluctuations. We analyze this effect
by assuming that string theory at energies $\sim H$ is
approximately local.  Therefore, by integrating
out
heavy degrees of freedom (of characteristic mass $M$),
we can write a local
effective action for the inflaton
field at momentum scale $H$.
We identify which terms contribute the largest
effect for large $M$
(the leading irrelevant operators)  and recover the basic
$H^2/M^2$ estimate for the imprint on the CMBR.
 We then show that all weakly coupled string models, and in
fact all ordinary field theoretic models in the
absence of fine tuning, give unobservably
small effects.

In Section 4, we turn to strongly coupled string
theory in a search for lower fundamental mass
scales which may lead to larger effects.  We analyze M theory models
of the Ho\v rava-Witten type using the phenomenologically
appealing grand unified compactifications discussed
in \witstrong .  We show these models give effects of
size $\leq 10^{-7}$, too small to be observed,
but larger than the weakly coupled string models because
the fundamental eleven dimensional Planck
scale here is lower,  $\sim 5 \times 10^{16}$ GeV.
We go on to discuss $G_2$ compactifications of M theory.
Here, rather than having, roughly speaking, one
large dimension as in the Ho\v rava-Witten
case, we can have four large dimensions, as the singularities
supporting gauge dynamics are codimension four
\refs{\gtwofiber,\gtwochiral}.
If we abandon the requirement
of precision grand unification and allow our compactification
manifold to get as large as possible, while remaining
consistent with the four dimensional character
of inflation, we can make the imprint on
the CMBR order one,
and hence potentially observable.   In these models the fundamental
eleven dimensional  Planck mass is $m_{11} \sim H \sim 7 \times 10^{13}$ GeV.
We also consider the early cosmology of models with low unification
scale $m_* \sim $ TeV. In these models the size of extra dimensions varies in the
course of cosmological evolution, but the size of the imprints of the
type we consider remains small.

In Section 5,  we discuss in detail the requirements
necessary to observe these effects and distinguish them
from other phenomena.   It turns out that corrections of
this type over the range of wavelengths accessible in scalar CMBR
observations look like a change in the power law, or ``tilt"
of the observed power. Such a change can
be mimicked by a change in the inflationary potential.
What cannot be mimicked is the
differential effect in the scalar and tensor
fluctuations due to short distance physics.
This point was first made in \conscon .
Ordinary inflationary fluctuations, without new physics,
obey  ``inflationary consistency conditions"
connecting scalar and tensor quantities.
New heavy physics predicts a violation of these conditions \conscon .
This is the unambiguous signal of new physics.

We then show that the differencing inherent in the
inflationary consistency conditions means
that the actual
 signal is not of the size $\sim r$ as discussed
above, but is further suppressed by
what is called an inflationary ``slow roll parameter''
which can range from $\sim .001-.06$  in various
models.  So the size of the measurable effect is somewhat
smaller than initial estimates suggest.

Finally, we point out that this unambiguous signal is
very challenging to measure.   First, it not only
requires precision data for the scalar fluctuations,
which are rapidly accumulating, but it also
requires precision data for the tensor fluctuations,
which have not even been observed yet.  Forthcoming
experiments may however be able to observe the tensor
fluctuations if inflation occurred at a  high
scale by observing the B-mode polarization
component of the CMBR.
We argue that cosmic variance limited measurements  over a substantial range
in wavenumber of this quantity will be necessary to detect these signals.
This is a formidable experimental challenge.

In Section 6, we conclude.

\newsec{Slow Roll Inflation}

We begin with a review of the basic tenets of inflation.
We will parameterize the
inflationary potential $V$ by a scale $M^4$ and a dimensionless function
${\cal V}$; $V = M^4 \cal V$.
We will work in a spatially flat FRW universe
with the metric
\eqn\metric{ ds^2 = -dt^2 + a^2(t) d\vec x^2.}
The independent background
field equations then reduce to
\eqn\backeqs{\eqalign{
& 3H^2 = {1 \over m^2_4} \Bigl[ {\dot \phi^2 \over 2} + M^4 {\cal V} \Bigr] \cr
& \ddot \phi + 3H \dot \phi + M^4 {\partial {\cal V} \over \partial \phi} = 0,
}}
where $H = \dot a/a$ is the Hubble parameter, $m_4 \sim 2.4 \times 10^{18} {\rm GeV}$
is the reduced four dimensional Planck
mass,  and dots denote time derivatives. The
main feature of inflationary dynamics in the slow roll approximation
is that we can ignore the acceleration of the
scalar field, because the cosmological expansion
has the effect of friction and nearly freezes the scalar
on the potential slope.
The universe is dominated by the scalar field
potential energy and undergoes a period of
rapid expansion. The usual parameters which characterize the
validity of the slow roll approximation
are
\eqn\slowrolpar{\eqalign{
& \eta = {\ddot \phi \over H \dot \phi} \cr
& \epsilon = {3 \dot \phi^2 \over 2 M^4 {\cal V}}.
}}
The slow roll approximation is then formally defined as the regime
$| \eta | , | \epsilon | \ll 1$.
The relative importance of these
parameters depends on the model of inflation.
For example, as we will see below in the case of natural inflation \naturalinfl,
or modular inflation, \modular, $\epsilon \ll \eta$.
Thus the deviations from  slow roll are
mainly coded in the parameter $\eta$. In contrast,
in the simplest model of chaotic inflation driven by a mass term,
$\epsilon \sim m^2/H^2 >> \eta = 0$ in the slow roll regime.

In the slow roll approximation, the equations \backeqs\ become
\eqn\slowreqs{\eqalign{
& 3H^2 = {M^4 \over m^2_4}  {\cal V} \cr
&3H \dot \phi + M^4 {\partial {\cal V} \over \partial \phi} = 0.
}}
Using these equations, one readily finds the slow roll parameters
in terms of the potential function ${\cal V}$:
\eqn\slowrolparpot{\eqalign{
& \eta = \epsilon - m^2_4 {\partial^2_\phi {\cal V} \over {\cal V}} \cr
& \epsilon = m^2_4 {[\partial_\phi {\cal V}]^2 \over 2{\cal V}^2}.
}}
The equations \slowreqs\ can now be integrated;
they yield
\eqn\slowrsoln{
{da \over a} = - {1\over m^2_4} {{\cal V} \over
\partial_\phi {\cal V} } d\phi , }
which separates variables for any potential ${\cal V}$. The
solution is
\eqn\infsolut{
a \simeq a_0 \exp\Bigl[{1\over m^2_4} \int^{\phi_0}_\phi
d\phi {{\cal V} \over \partial_\phi {\cal V}} \Bigr]
 \simeq a_0 \exp\Bigl[ {{\cal V}_0 \left( \phi_0 - \phi \right)
\over m^2_4 \partial_\phi {\cal V}_0} + ... \Bigr]}
in the slow roll regime. Hence, the universe will undergo
rapid expansion while the vev of the inflaton may change
only minutely. The space-time geometry is
approximated by a future half of de Sitter space during this
period. Eventually however the change of the inflaton vev
accumulates enough for the inflaton to depart the slow roll
regime, and the potential becomes steeper. The inflaton
approaches the minimum of the potential, begins to oscillate
around it and produce matter particles,
reheating the inflated universe back to
temperatures which will eventually produce
the universe we inhabit.

Let us imagine that at late times the
vacuum energy vanishes and inflation terminates such that there are no
cosmological event horizons. This avoids conceptual difficulties with quantum
gravity in spacetimes with cosmological horizons, but
suffices to illustrate the
main features of inflationary dynamics in the spacetime language.
The causal structure of the universe is then
given by the Penrose diagram of Fig. 1.

\ifig\causal{Causal diagram of an
inflationary model. The dashed past null line is the
true particle horizon, but it could also be a null singularity. The
future null line is the future infinity.
The shaded area denotes the region of exit from inflation and reheating.
The thin solid line is a worldline of any
spacelike separated object from an observer at the center of the
space. The bold solid line is the apparent horizon. Its shape is
characteristic of inflation in the past, and radiation domination
followed by matter domination in the future.}
{\epsfxsize2.0in\epsfbox{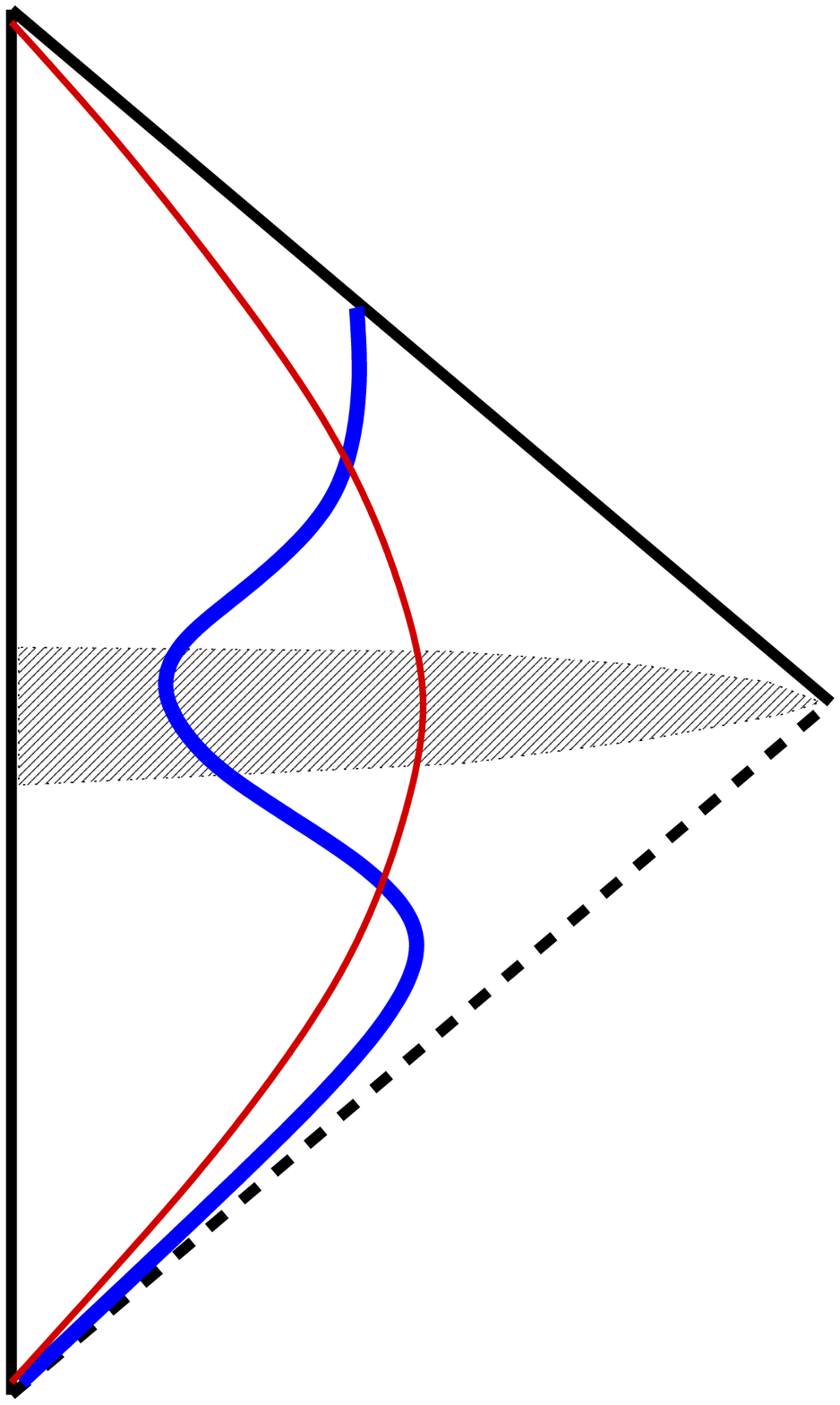}}

In the diagram, the region of geometry below the particle horizon
(dashed line) is irrelevant for the future evolution
as long as the period of inflation was sufficiently long. The
future infinity appears as a consequence of our
requirement for global exit from inflation.
The spacetime below the reheating regime is the inflationary region,
while that above it is the postinflationary, decelerating FRW
universe. The thin solid line denotes any
object spacelike separated from us, and for example
is the worldline a distant galaxy follows after it forms.
The bold solid line represents the apparent horizon, which plays
central role for controlling the dynamics of inflation, as we will
now elaborate.
During inflation, it
starts out almost null and ``outward" directed, and then it flips
``inward". This reflects that $H \sim$ const.
during inflation.
It ensures that the apparent horizon plays the
role of the causal censor, limiting the amount of information
which can fit inside an inflating region. The
spacetime will therefore obey cosmic no-hair theorem, and inflation
will succeed in getting rid of initial inhomogeneities. This may be viewed as
another example of the cosmological variant of the holographic
principle \fs\rb. The structure of the spacetime is fully coded
on the preferred screen, i.e. the apparent horizon. Its
area is small during inflation because $H$ must be large,
and hence the Hubble region is censored from excessive outside
influence, because only a limited amount of information
can fit in the interior.
Moreover, most of the objects inside the Hubble region
are in the thermal bath of fluctuations located in the region when
the apparent horizon is almost null \hks, with the cosmological
Hawking temperature $T_H = H/2\pi$.
Since the inflaton is much lighter than
the Hubble scale during inflation, the interactions with the thermal
quanta cause its vev and the background metric
to fluctuate.

Because of these quantum fluctuations, the
inflaton is not exactly frozen to its slowly varying
background vev. Instead it hops on the
potential around the background value.
Thus inside of some regions of the universe inflation
may terminate a little later, because quantum effects push
the inflaton a little farther up the plateau. These regions end up
a fraction denser than their surroundings, and the matter in them
begins to condense sooner, attracting additional matter from the neighborhood
and eventually forming clusters and galaxies due to the classical
Jeans instability. The
fluctuations therefore induce small inhomogeneities on the perfectly smooth geometry
left by inflation, which is measured experimentally via its imprint on the cosmic
microwave background radiation,
${\delta \rho/ \rho} \sim  \delta T/T$, thanks to the Sachs-Wolfe
effect. This is directly measured
by the COBE \cobe\ satellite, and by the BOOMERanG \boomerang\ and MAXIMA \maxima\
experiments, which set the normalization for the
inhomogeneities at around $\delta \rho/\rho \sim 10^{-5}$.
They further observe that the spectrum of
inhomogeneities is nearly scale-independent.

To determine the imprint of the fluctuations
quantitatively we can use perturbation
theory.
In perturbation theory the fluctuations can be decomposed with respect
to their transformation properties relative to the residual diffeomorphisms into
scalar, vector and tensor modes. The vector modes decouple during
inflation. Thus only the scalar and tensor modes are
produced. The scalar modes cause
density (and therefore CMBR temperature) fluctuations. The tensor
modes are primordial gravitational waves produced by inflation, and affect the
polarization of CMBR.

A heuristic derivation of the
scalar density contrast is as follows: the
RMS fluctuation of the field induced by the thermal fluctuations
is $\delta \phi = \dot \phi \delta \tau$, and that of energy density is
$\delta \rho = C \rho H \delta \tau$, where
$C$ is a numerical coefficient of order unity,
whose precise value depends on the details of postinflationary cosmology.
The function $\delta \tau$ is the time delay imprinted by the fluctuations
on the vev in different regions of space.
Combining these equations, one finds
\eqn\dencont{{\delta \rho \over \rho} = C { H \over \dot \phi} \delta \phi}
and then one needs to compute the RMS fluctuation of the inflaton $\delta \phi$.
As we will discuss in more detail below, fluctuations of the
transverse traceless modes of the
graviton (which obey free scalar field equations)
also contribute to the density variations.

In order to determine precisely how the quantum fluctuations
of these fields evolve into temperature anisotropies
in the sky today, one must first compute their effect
on the curvature, and then use gauge-invariant gravitational
perturbation theory to evolve the perturbation forward to
the present era.
One can define a gauge-invariant variable $\zeta$, which is
well approximated by the right hand side of \dencont\ as modes exit from the de Sitter
horizon during inflation, and which
is approximately constant between this time,
and when the mode re-enters the cosmological horizon later.
 At this later time $\zeta$ is well
approximated by $\delta \rho / \rho$, establishing \dencont\ \bst.
This stage in the process is purely classical, because
energy scales below $H$ correspond to scales outside
the causal horizon, and so coherent quantum
fluctuations do not contribute at these wavelengths.
The correct procedure is therefore to compute the quantum fluctuation of the
inflaton field in de Sitter space, and then use it to evaluate $\zeta$
at the time the fluctuation exits the horizon; i.e., at momentum
$p = H$.

As pointed out in \wang,
if the slow roll approximation breaks down this procedure will not
be accurate (however, see \cgkl ).
For the sake of simplicity we will restrict
ourselves to models where this is not a concern.

To compute the quantum fluctuation itself,
one treats the fluctuating field as a perturbation
around the de Sitter background and
computes the mean-square variance as the
(appropriately normalized) Fourier component
of an equal-time two-point function evaluated at 3-momentum $p = H$:
\eqn\inffluc{
(\delta \phi)^2 \sim \langle \phi(p) \phi(-p) \rangle |_{p=H},
}
where $\phi$ represents either the inflaton
or a physical mode of the graviton.  The normalization
is determined by the more detailed computation we perform
below.  In standard
inflation, this is done assuming the inflaton
is a free, minimally coupled scalar.  As we will
demonstrate in section 3, interactions with massive particles
will modify the 2-point function and affect
the spectrum of fluctuations.  As long as the self-interactions
of the inflaton (either in the classical potential,
or induced by quantum corrections)
are weak at energy scale $H$, so that a perturbative
expansion is valid at this scale, this procedure is well defined.
Of course, more general theories will involve strong
coupling, but generically will also violate the
observed constraints on $\delta \rho/\rho$.

Before considering such complications, we review the standard calculation.
We begin by approximating the geometry by
a future portion of de Sitter space.  With $a = a_0 \exp(Ht)$ in
\metric, the inflaton
field equation is
\eqn\eqmot{
\ddot{\phi} + 3 H \dot{\phi} - e^{-2 H t} \vec{\partial}^2 \phi  =
H^2 \Bigl(\eta^2 \partial_\eta^2 \phi - 2
\eta \partial_\eta \phi - \eta^2 \vec{\partial}^2 \phi \Bigr) = 0,}
where we have transformed to the conformal
time $\eta \equiv - H^{-1} e^{-H t}$.  We can quantize $\phi$
by considering the
general solution
\eqn\sol{
\phi_p(\eta) = {\sqrt{\pi}  \over 2} H \eta^{3/2} \left[A_k
H^{(1)}_{3/2} (k \eta) + B_k H^{(2)}_{3/2}(k \eta ) \right].}
Choosing the vacuum which matches the flat space case in
the infinite past $\eta \rightarrow \infty$ and in the high
frequency limit $k \rightarrow \infty$, we find that positive
frequency modes are $A_k = 0$, $B_k = -1$.
Then the mode expansion in Minkowski space is
\eqn\field{ \phi(\vec{x}, t) = (2 \pi)^{-3/2} \int
d^3k \left[ a^{\dagger}_k \phi_k(t) e^{i \vec{k} \cdot \vec{x}} +
a_k \phi^*_k(t) e^{-i \vec{k} \cdot \vec{x}}  \right] }
where
\eqn\fourier{ \phi_k(t) = {i H \over k \sqrt{2 k}} \left(1 + {k
\over i H} e^{-H t} \right) \exp{\left({ik \over H} e^{-H t
}\right)}.}
The positive frequency 2-point function is
\eqn\wightp{
\eqalign{
G^+(x, x') &\equiv \langle 0 |  \phi(x) \phi(x') | 0 \rangle
= {1 \over (2 \pi)^3}  \int d^3 k \, \times \cr
& \times
e^{-i\vec{k} (\vec{x}-\vec{x}')}
 \left[ {H^2 \over 2 k^3} +
{e^{-H(t + t')} \over 2 k} + {i H \over 2 k^2} \left( e^{-Ht} -
e^{-H t'} \right) \right] \exp \left( -{i k \over H}
\left( e^{-H t} - e^{-H t'} \right) \right).}}
To evaluate the fluctuations of the inflaton at lowest order,
we compute the quantity
\eqn\fluc{
\langle \phi(x) \phi(x) \rangle = {1 \over (2 \pi)^3}  \int d^3 k
 \left(  {e^{-2 H t} \over 2 k}+
 {H^2 \over 2 k^3}\right) = {1 \over (2 \pi)^3} \int {d^3 p \over p}
\left({1\over 2} + { H^2 \over 2 p^2} \right),
}
where $p = e^{-Ht} k$ is the physical momentum conjugate
to the proper distance $\tilde{x} = e^{H t} x$.

The first term, which gives a UV-divergent contribution, is identical
to the flat space result and should therefore be ignored.  In other words,
we are interested only in effects proportional to $H$, not in
flat-space fluctuations which can be renormalized away.
The second term is peculiar to de Sitter space and requires more careful treatment.

The magnitude of the  fluctuations is determined by their power $P_{\phi}(k)$, defined by
$\langle \phi(x)^2 \rangle = \int {dk\over k} P_\phi(k)$.
Then the mean-square spectrum of fluctuations is
$ \langle |\delta \phi|^2 \rangle = P_\phi(H)$.
{}{}From \fluc\ , we see that
\eqn\flucb{
\langle \phi(x) \phi(x) \rangle = {1 \over 2 \pi^2} \int {d k \over k}
\left({k^2 \over 2} + { H^2 \over 2}\right),}
so, neglecting the first term as explained above, we obtain
\eqn\meansquare{
\langle |\delta \phi|^2 \rangle =
 {{H^2 } \over 4 \pi^2}.}
This gives $\delta \phi = H/2\pi$,
finally yielding
\eqn\dencontf{ {\delta \rho \over \rho} = {C \over 2\pi} { H^2 \over \dot \phi}. }

At this point, it is clear how to incorporate interactions
into the calculation.  If the theory contains a massive
field (with mass $M \gg H$) which couples to the inflation,
we can integrate it out using standard field theory techniques
and obtain an effective potential for the inflaton.  As can be seen
from the two point function \wightp, such a procedure yields--in
addition to the ordinary flat space terms--terms proportional
to $H^2/M^2$, $p^2 H^2/M^4$, etc. It is important to note that
no cutoff or Planck scale comes into these corrections.
The highest probe energy available in inflation and later visible in the CMB is $H$.
As discussed in Sec. 3, it is
these contributions we are primarily concerned with in this paper.

We can re-express \dencontf\ in terms of the inflationary
potential using the slow roll equations \slowreqs. It is
\eqn\dencontfpot{ {\delta \rho \over \rho} = {C \over 2 \sqrt{3} \pi}
{M^2 \over m^3_4} { {\cal V}^{3/2} \over \partial_\phi {\cal V}}. }
This is the familiar formula for scalar fluctuations in inflation.
We note that the so-called scalar power spectrum $\delta_S^2$ is
related to the density contrast by $\delta^2_S = (2/5C)^2(\delta
\rho/\rho)^2$, and using \dencontfpot\ we can express it as
\eqn\scaspect{ \delta^2_S = {1 \over 75 \pi^2}
{M^4 \over m^6_{4}} { {\cal V}^{3} \over [\partial_\phi {\cal V}]^2} }

The causal structure of the inflationary spacetime depicted in Fig. 1.
provides a straightforward understanding of
the emergence of a (nearly) scale-invariant spectrum of
fluctuations. A quantum fluctuation which seeds a galaxy
is created just before its worldline intersects the
apparent horizon. At that instant, it is as big as the Hubble horizon.
Then it is expelled outside of the apparent
horizon, where it freezes, and remains frozen until horizon
reentry in distant future. After reentry the fluctuation becomes
dynamical and evolves as dictated by gravitational instability.
Scale invariance then follows from causal evolution if $H \simeq$
const., because the fluctuations of very different wavelengths
are produced with the same amplitude. The evolution of the
fluctuations can initially be described well by linear
perturbation theory. However,  nonlinearities eventually develop
because of nontrivial interactions with the environment.
In the matter dominated era, the fluctuations evolve differently
before decoupling than after it. Before
decoupling, the universe is opaque and therefore the baryonic matter
is influenced by radiation pressure, which competes with
gravitational collapse. This results in the emergence of
acoustic oscillations, with characteristic peaks
imprinted on the CMBR. The peaks appear
because the perturbations whose wavelengths are half-integer divisors of
the sound horizon (i.e. the largest distance sound can travel
within the time of recombination)
at decoupling can complete full oscillation
cycles.  The location and the heights of the peaks measure very
accurately the cosmological parameters, in particular the Hubble
parameter at decoupling.

Before turning to the specifics of modular inflation,
we briefly review the mechanism for generating
tensor fluctuations during inflation.
These are just the gravitational waves, and correspond to the
transverse-traceless metric fluctuations $h_{kl}$.
They obey the linearized field equation $\nabla^2
h^k{}_l = 0$, where the covariant derivatives and raising and
lowering of indices is defined relative to the background metric
$g_{\mu\nu} = {\rm diag} (-1, a^2(t) \delta_{kl})$. Therefore each
of the two graviton polarizations obeys the free massless scalar
equation, and it is straightforward to quantize them in de Sitter
space, in precisely the same way as in eqs. \eqmot\ -
\meansquare. In particular the root mean square fluctuation of the
graviton is $\langle \delta h_{kl}
\rangle \simeq H/2\pi$. However the formula for
the tensor power spectrum is different than for the scalar. It is
directly proportional to the fluctuation of the metric,
\eqn\tenpow{\delta_T^2 = {1 \over 2\pi^2 } {H^2 \over m^2_4} =
{1 \over 6 \pi^2 } { M^4 \over m^4_{4}} {\cal V} }
by slow roll equations \slowreqs.
The tensor nature of these fluctuations induce
oscillations in the plasma during decoupling  which
polarize the CMB photons in an observable way \huandwhite.

The ratio ${\cal R} = \delta^2_T/\delta^2_S$ is a characteristic of the
inflationary model, and is given by
\eqn\ratio{{\cal R} = {25 \over 2} {m^2_4 [\partial_\phi
{\cal V}]^2 \over {\cal V}^2}}
It is straightforward to verify that in terms of the slow roll
parameters, ${\cal R}$ is given as
\eqn\ratioslr{{\cal R} = 25 \epsilon\ .}

The fluctuation spectra produced in inflation
are not exactly scale-invariant.
If the background inflaton vev were exactly frozen, and
the geometry precisely de Sitter,
the prediction for fluctuations \dencontf\tenpow\ would have been
time-independent, and therefore exactly scale-invariant.
In reality,  there is weak time-dependence in \dencontf\
because the inflaton
is sliding down the plateau. This time dependence, manifest in the
variation of $H$ and $\dot \phi$,
translates into scale dependence of fluctuations, and produces a spectrum
which is not exactly scale-invariant.
This departure from scale invariance is a function of
the specifics of inflationary model as defined by the
potential. Below we will consider the details in the case of
modular inflation.

\subsec{The Specifics of Modular Inflation}

To proceed, we need to determine more closely the form of the
inflaton potential.
For definiteness, we approximate here the potential function ${\cal V}$
by an inverted parabola
\eqn\potspec{ {\cal V} = 1 - \Bigl({\phi \over \mu} \Bigr)^2}
This approximation is generically valid
in at least some region of the primordial
universe which begins to inflate, if the inflaton is a modulus.
The modulus begins near the maximum
of the potential. Then the slow roll conditions yield
\eqn\slowroleqs{\eqalign{
& H = {M^2 \over \sqrt{3} m_4}
\sqrt{1- ({\phi \over \mu})^2}  \cr
& \dot \phi = {2 M^2 m_4 \over \sqrt{3} \mu^2}
{\phi \over \sqrt{1- ({\phi \over \mu})^2}}
}}
The slow roll parameters for \potspec\ are initially
\eqn\srpexp{\eqalign{
&\eta \simeq \epsilon - m^2_4 {\partial^2_\phi {\cal V} \over
{\cal V}} \simeq
{2 m^2_4 \over \mu^2} \cr
&\epsilon \simeq {m^2_4 \over 2} {[\partial_\phi {\cal V}]^2 \over {\cal V}^2}
\simeq {2 m^2_4 \phi^2 \over \mu^4} < \eta
}}
and hence, the parameter $1/\mu^{2}$ which we introduced in the potential
\potspec\ is equal to $1/m_4^{2}$ multiplied by a small parameter $\eta /2$.
This guarantees that the potential is sufficiently flat
to support inflation.

We could now solve these equations directly. However, rather than
integrating to find the time-dependence, it is more instructive to
solve the equation \slowrsoln. We will use the number of efolds
before the end of inflation, or equivalently, the value of the
scale factor $a$, as the cosmic clock. First, we define the number
of efolds that universe has expanded by to be \eqn\efolds{ N =
\ln\Bigl({ a \over a_0} \Bigr) = \int^t_{t_0} dt H =
\int^\phi_{\phi_0} d \phi {H \over \dot \phi} } Then using
\slowroleqs\ we can explicitly integrate this to find \eqn\neqn{N
= {1 \over \eta} \Bigl[\ln\Bigl({\phi \over \phi_0} \Bigr) + {1
\over 2\mu^2} \Bigl(\phi_0^2 - \phi^2 \Bigr) \Bigr]} Here $\phi_0$
is the initial value of the inflaton. In modular inflation,
$\phi_0$ would typically be near the top of the potential, in this
case near zero. Such initial conditions produce a huge amount
of inflation, as is clear from \efolds, which diverges in the
limit $\phi_0 \rightarrow 0$. Of all that expansion, we can
only observe the final $60$
efolds or so, during which the universe expands by a factor of
about $10^{26}$. Any indications of expansion beyond that would be
completely outside of the current size of the universe, and hence
not accessible to our observations. Because we are only interested
in the last $60$ efolds, we can take $\phi_0$ to be near
its value at the end of inflation. Inflation ends when the
slow-roll conditions cease to be valid, i.e. when $\eta,
\epsilon \sim 1$. This occurs when the vev of $\phi$ grows to
$\partial^2_\phi V \sim H^2$, which in the case of modular
potential \potspec\ happens when $\phi \sim \mu$. Because a small
change in the value of $\phi$, of order $\sim e$, yields $N \sim
\eta^{-1}$ efolds of inflation, we require that $\eta \sim
1/70$, which is enough to solve the horizon and flatness problems.
In that case the latter two terms in eq. \neqn\ are essentially
negligible compared to the logarithm, and we will drop them
hereafter. We note that in this case the other slow roll parameter
is $\epsilon \simeq 2m^2_4/\mu^2 e^2 \simeq \eta/e^2$.

We now define ${\cal N} = N_* - N$ as the number of efolds left
before the end of inflation. This variable is convenient to make
contact with large scale structure and CMB observations. In terms
of it, we can write down the solutions as \eqn\solns{\eqalign{
&{\cal N} = {1 \over \eta} \Bigl[ \ln\Bigl({\mu \over
\phi}\Bigr) + {\phi^2 - \mu^2 \over 2 \mu^2} \Bigr] \cr &H = {M^2
\over\sqrt{3} m_4} \sqrt{1- ({\phi \over \mu})^2} \cr &a =
a_{final} e^{-{\cal N}} }} Inflation now lasts from when ${\cal N}
\sim 60$ or larger, to about ${\cal N} = 0$, at which point the
higher-order terms in the modular potential, ignored for clarity
in \potspec, become important and reverse the sign of the
effective mass term of $\phi$.

Because the rolling of the scalar down the potential is slow, the
Hubble parameter and the scalar field change very little, and
hence the amplitude of fluctuations remains nearly constant
throughout inflation. Therefore the fluctuations are being
incessantly produced with an almost constant value, and deployed
outside of the horizon. They stay there until a long time into the
future, when the Hubble horizon eventually grows large enough, and
they cross back inside, and start to collapse. These are the
fluctuations we observe on the sky. The weak time dependence of
$H$ and $\dot \phi$ implies that $\delta \rho/\rho$ is weakly
scale-dependent. We trade the time dependence off for the scale
dependence by the horizon crossing matching, defining the comoving
momentum $k$ of the fluctuation at horizon-crossing by
\eqn\horcross{ k = aH} For modular inflation solution \solns\ this
enables us to explicitly evaluate \dencontf\ as a function of $k$.
First, we note that \eqn\kmatch{ k \simeq k_0 e^{-{\cal N}}} where
$k_0$ is the comoving momentum leaving the Hubble horizon at the
end of inflation. In terms of it, we find \eqn\denconmom{ {\delta
\rho(k) \over \rho} \simeq {2 C \mu M^2 \over 8 \pi \sqrt{3}
m^3_{4}} \Bigl( {k_0 \over k} \Bigr)^\eta \Bigl[1 - \Bigl( {k
\over k_0} \Bigr)^{2\eta} \Bigr]^{3/2} } At $50$ efolds before the
end of inflation, the COBE measurements set the overall
normalization to $\delta \rho/\rho |_{50} \sim 5 \times 10^{-5}$.
Thus $C \mu M^2/m^3_{4} \simeq 8 \times 10^{-4}$. Taking now $C =
{\cal O}(1)$ and $\mu \simeq \sqrt{140} m_4$ \srpexp, we obtain
\eqn\scaleinf{ {M \over m_4} \simeq 8.2 \times 10^{-3} } In this
case, the Hubble scale at inflation is, using the first of
\slowroleqs,
\eqn\hsnum{H \simeq 5.2 \times 10^{13} {\rm GeV}\ ,}
which is within the bound allowed by COBE and
large-scale structure measurements, c.f. \kamkos.

The scale dependence is
conveniently represented by defining the spectral index $n_S$ (or the tilt)
and its gradient as
\eqn\tilt{\eqalign{
& n_S = 1 + 2 {d \ln {\delta \rho \over \rho} \over d \ln k} \cr
& \nu_S = { d n_S \over d \ln k}  =
2 {d^2 \ln {\delta \rho \over \rho} \over d \ln k^2} }}
For the modular  inflation model we find
\eqn\tiltresult{
n_S = 1 - 2 \eta -  {6 \eta \over ({k_0 / k})^{2\eta} - 1} .}
Numerically, using \tiltresult, we find the tilt (at the scales corresponding to $50$
efolds before the end of inflation) to be
$n_S \simeq 0.95$ and $\nu_S \simeq - 24 \times 10^{-3}$.

The tensor power spectrum is found by substituting \solns\ and
\kmatch\ into \tenpow. We find
\eqn\tenpowmod{\delta^2_T = {1 \over 6 \pi^2} {M^4 \over m^4_{4}}
\Bigl[1 - \Bigl({k \over k_0} \Bigr)^{2\eta} \Bigr]. }
Therefore, the ratio of tensor to scalar spectrum of
fluctuations is, using $\epsilon = {\eta \over e^2}$, precisely
\eqn\ratmodul{ {\cal R} = 25 \epsilon,}
which numerically is ${\cal R} \simeq 4.8 \times 10^{-2}$.

Above we have focused on the most familiar case, where the departure
from the slow roll regime is dominated by the quadratic terms
in the potential.
It may however happen that the inflaton mass scales are
smaller than the scales set by the vev, such that the termination
of inflationary conditions is controlled by
higher polynomial contribution to the inflaton potential.
The numerical values we have obtained
for \scaleinf\ and \hsnum\ clearly are sensitive to the precise form
of the inflationary potential during the last 60 efolds, and it is of interest
to determine the range of these parameters. To do so, one can parameterize
different modular inflationary models by
the potential function
\eqn\pothighpol{
{\cal V} = 1 - \Bigl({\phi \over \mu}\Bigr)^n ,}
which yields, in the slow roll regime, the field equations
\eqn\highslow{\eqalign{
& H = {M^2 \over \sqrt{3} m_4} \sqrt{1- \Bigl({\phi \over \mu}\Bigr)^n }\cr
& \dot \phi = {n M^2 m_4 \over \sqrt{3} \mu^n}
{\phi^{n-1} \over \sqrt{1 -\Bigl({\phi \over \mu} \Bigr)^n}}.
}}
It is straightforward to find the solution,
\eqn\solpol{\eqalign{
& a = a_{final} e^{-{\cal N}} \cr
&{\cal N} \simeq {\mu^n \over n(n-2) m^2_4 \phi^{n-2}}
+ {(n-4)\mu^2 \over 2n(n-2) m^2_4}, }}
and determine the density contrast. It is
\eqn\dencontpoly{
{\delta \rho \over \rho} = {C \over 2n\pi \sqrt{3}}
\Bigl({M \over m_4 } \Bigr)^2 {\phi \over m_4}
\Bigl({\mu \over \phi} \Bigr)^2 \Bigl[1-\Bigl({M \over m_4 } \Bigr)^n
\Bigr]^{3/2}.}
{}From the COBE normalization $\delta \rho/\rho|_{50} \sim 5 \times
10^{-5}$ we can derive an estimate of the scale of inflation. While
there is some sensitivity to the initial condition, we find
$M  \sim 4 \times 10^{-2} n^{1/2} (n-2)^{1/4} m_4$, or
\eqn\modpolpoth{
H \sim {\rm few} \times 10^{14} n \sqrt{n-2} ~{\rm  GeV}.}
But in light of the bound $H < 7 \times 10^{13}$ GeV on
the Hubble scale of inflation
\kamkos, we see that the modular inflation
models with $n>2$ are excluded already,
and we can ignore them henceforth.

In some cases, most of the late inflationary expansion
can occur during the final approach of the inflaton to the minimum
of the potential; this the the scenario of chaotic inflation
\linde. In these cases, the potential is
\eqn\potfuncch{
V = \frac{\lambda}{n} \phi^n,}
where $V$ is the dimensionful quantity $V = M^4 {\cal V}$.
In the slow roll approximation the field equations reduce to
\eqn\chaoslow{\eqalign{
& H = \sqrt{\lambda \over 3n} {\phi^{n/2} \over m_4}\cr
& \dot \phi = -\sqrt{n \lambda \over 3} m_4 \phi^{n/2-1}.
}}
The slow-roll solution is
\eqn\chaosoln{\eqalign{
& a = a_{final} e^{-{\cal N}} \cr
&{\cal N} \simeq {1 \over 2n} \Bigl({\phi \over m_4}\Bigr)^2. }}
The density contrast is
\eqn\chaodens{
{\delta \rho \over \rho} \simeq
{\sqrt{\lambda} \over 2 \pi \sqrt{3} n^{3/2}} {\phi^{n/2+1} \over m^3_4},}
and using the COBE normalization we can straightforwardly
determine $H$ during inflation. It is
\eqn\hchao{
H \simeq 4\pi \times 10^{-6} \sqrt{n} ~ m_4,}
which is a factor of $\sqrt{n/2}$ higher than the corresponding
value in the case of quadratic (sub)leading potential.
It is straightforward to determine the spectral index for scalar perturbations. It is
\eqn\scaltiltchao{
n_S = 1 - \frac{n+2}{2\CN}\ .
}
Hence, in general, chaotic inflationary models driven by higher polynomial
terms tend to yield a higher value of $H$ during inflation, but they also
give steeper potentials and therefore will yield larger values
of the spectral index. The slow-roll parameters for chaotic inflation are
\eqn\slowrolcha{\eqalign{
& \eta \simeq  -{n (n-2) \over 2} ~{m^2_4 \over \phi^2} \cr
& \epsilon \simeq {n^2 \over 2} ~{m^2_4 \over \phi^2} \sim \eta.
}}
The ratio of tensor to scalar perturbation power obeys \ratio, ${\cal R} = 25 \epsilon$,
by virtue of \scaspect,\tenpow\ and \slowrolcha. For low powers $n$, the
parameter $\epsilon$ now determines the duration of inflation, which therefore
means that the ratio ${\cal R}$ is only weakly sensitive to the specifics of the
potential, giving similar tensor power for different forms of $V$.

\newsec{Imprint of Heavy States}

We now turn to the heart of our work--finding the imprint of new,
heavy physics on the fluctuations
discussed in the previous section.    We will assume that the scale of
inflation $H$ is much smaller
than the Planck mass,   $H \ll m_4$, so that a field theoretic
treatment of gravity is appropriate.
Further, we will assume that the mass scale of new physics
$M$ is much larger than $H$, $M  \gg H$,
and then assume that we can represent the
effects of this new physics at the scale $H$
by ``integrating it out" and writing  an effective
field theory for the inflaton field.   These
assumptions--which rely on low energy locality and renormalization
group ideas--are obviously correct in a field theoretic
context,  are obviously correct within string perturbation
theory around supersymmetric vacua,
and are correct in the known nonperturbative definitions of
string and M theory in supersymmetric
backgrounds.  The enduring mystery of the cosmological constant,
and the associated mysteries of string theory in de Sitter
space, make these assumptions plausible, but
not ironclad, in the present context.   We make them anyway.\foot{For an example of
a speculation on how locality might break down in
de Sitter space string theory see \bankscosmo .}

We then can encode all the new physics by writing an
effective field theory for $\phi$ at the scale
$H$.  The scale $H$ is appropriate since,
as we see in \inffluc,  that is where we
evaluate inflaton correlation functions to compute
the size of ${\delta \rho / \rho}$.

Instead of writing a fully covariant effective action for $\phi$,
let us just note that the curvature
of de Sitter space is proportional to $H^2$, and so we use
this as an additional dimensionful
parameter in constructing terms.     The interactions of the
inflaton must always be very weak to give phenomenologically acceptable
values of ${\delta \rho / \rho}$.
This is usually enforced in specific models by some
combination of fine tuning, dynamics and  supersymmetry (broken at scale $H$).
So we will ignore inflaton
interactions.
Given these considerations
the most general Euclidean local action one can write down is of the form
(we have assumed $p \gg H$ and used flat space notation for simplicity):

\eqn\effact{
\eqalign{
S_{eff}[\phi] &= \int d^4 p \,\, \phi(p) \phi(-p)
 \{ p^2/2 + H^2/2 +c_0 H^2 (H^2/M^2)
+c_1 p^2 (H^2/M^2) + c_2 p^4/M^2\cr
&+ c_3 p^4/M^2 (H^2/M^2) + c_4 p^6/M^4 + \ldots \} .} }

This structure follows from the fact that only even
powers of momenta are allowed, and that
the curvature is $\sim H^2$.  Therefore,
 no odd powers of $M$ can appear.

Information about new physics is contained in the
coefficients $c_i$ and in the scale $M$.
{}From \inffluc\ we see that measurements of
${\delta \rho / \rho}$ help determine the value
of $\langle \phi(p)\phi(-p) \rangle$  at $p=H$.
{}From \effact\ it follows that

\eqn\corrat{
\eqalign{
  \langle \phi(p)\phi(-p) \rangle|_{p=H} &= H^2
 + c_0 H^2(H^2/M^2)
 + c_1 H^2 (H^2/M^2)
 + c_2 H^4/M^2\cr  &+c_3 H^2 (H^2/M^2)^2 +c_4  H^2 (H^2/M^2)^2 + \ldots }}

 The large $M$ corrections
to $\langle \phi(p)\phi(-(p) \rangle |_{p=H}$  organize
themselves into a power series in the
dimensionless ratio $r=H^2/M^2$.  We have assumed
that this ratio is small, so the only
terms that are potentially observable
are the ones with coefficients $c_1$
and $c_2$.  The term with coefficient $c_0$ is just
a renormalization of the potential which we can ignore.

 On very general grounds
the effect of new physics, whether field theoretic,
string theoretic, M theoretic, etc. is
proportional to $r=H^2/M^2$.
The coefficients $c_i$
must be computed, however, and can be
much smaller than one,  giving effects much smaller
than the naive expectation.

Several groups \refs{\martinbrand,\Niem,\kempf,\egks,\egkss} have previously analyzed a
special case of this situation.   They have added
an irrelevant operator
to Einstein gravity and directly computed its
effect on inflationary fluctuations by
solving the linearized wave equations.
This requires specifying new boundary
conditions at high momentum on the higher order
differential equation.   These boundary
conditions are not determined by the model itself.
The authors in \kempf\ impose the constraint that the solutions rapidly relax to the ``adiabatic"
vacuum shortly after they are created.  They find imprints of size $r \sim  H^2/M^2$,
consistent with our general result. In \refs{\egks,\egkss} the authors
study the general boundary condition and then focus on a special, different\foot{
The authors in \kempf\ speculated that this boundary condition was the same as their adiabatic
condition.   The results of \egkss\ show this is not the case.}  boundary condition
that results in effects of size $\sim (r)^n, ~ n \simeq .5$. This effect
is inconsistent with our effective
action result and so presumably this
boundary condition  violates
locality in some way.
We should also note that an effect of this functional
form would imply a nonanalytic dependence on $g_s^2$ and on
$\alpha'$ which would signal the
breakdown of perturbation theory at weak coupling.
All in all it seems  likely to us that the special boundary condition
chosen in \refs{\egks,\egkss} is unphysical.
The subtleties mentioned at the beginning of this section
make it impossible to definitively rule out such a result, however.

The above illustrates the virtue of the effective action
approach we are using here.   Equation \inffluc\
shows that the relevant momentum scale for these
processes is $H$, not $m_4$.  If there is a large
hierarchy between these scales--which is the situation
we are envisioning--then there should be no reason
to consider Planckian dynamics at all, e.g.,
short distance boundary conditions,
in studying the fluctuation problem.
One simply encapsulates all the unknown short distance
physics in an effective action.
All the subtlety of choice of boundary condition is buried in the
assumption of the existence of a local effective action.
Given that our world appears to be local,
this seems an excellent assumption.

Perhaps we should phrase things in a more optimistic way.
If experiments detect imprints in the CMBR
of strength $r^{.5}$ as predicted in \refs{\egks,\egkss}
it would imply a breakdown of locality in low energy string
theory, which might be a crucial
clue in solving the cosmological constant problem!
Alternatively, such results could also indicate that physics other than
inflation may be responsible for the origin of structure in the universe.

We now turn to the evaluation of the parameters
in the effective action \effact\ in some specific
physical situations.  First imagine a heavy
fermion field $\psi$ coupled to the inflaton via a
Yukawa interaction $\lambda \phi {\bar \psi} \psi$ .
A one loop graph in de Sitter
space of $\psi$ particles clearly induces
interactions of the form in \effact\ .   These produce
effects in the propagator  \corrat\ of size
$\sim \lambda^2 H^2/M^2$ with $M = m_{\psi}$\ the fermion mass.
Typically $m_{\psi} \sim \lambda \langle \phi \rangle \sim \lambda m_4$.
(We ignore slow roll parameters here.)
So these effects are $\sim H^2/m^2_4 \sim 10^{-11}$ and hence unobservable.
This result is quite general.   A particle
renormalizably coupled to the inflaton will {\it generically}
have a mass $\sim \langle \phi \rangle \sim m_4$ and
so the virtual effects of this particle will be of order
$H^2/m_4^2$: unobservably small.\foot{We thank S. Thomas for pointing this out to us.}
Exceptions to this
result can occur if counterterms are fine tuned
to make the  particle masses unnaturally small.
Then the virtual effects can be very large and
certainly observable.   An extreme case of this
limit has been studied in \kolbresonant\ where a fermion
becomes massless for a certain value of the inflaton field vev.
When this vev is reached during the slow
roll, fermions are produced copiously, sharply reducing $\dot \phi$
and so, by \dencontf, creating
a sharp increase in ${\delta \rho / \rho}$ for a short time.
This translates into a sharp peak
in momentum space in the fluctuation spectrum.

These phenomena require an additional level of fine tuning on
top of any fine tuning required to
make the inflaton potential well behaved.

Next we turn to weakly coupled heterotic string theory models
of the ``traditional" type:  $g_s^2 \sim .1, ~ m_s \sim m_4 \sim  10^{19}$ GeV.
The Calabi-Yau compactification radii are
hence also of order $1/m_s$.
It will be useful to be able to vary this
scale independently, so we will denote it $1/m_{CY}$.
We do not understand inflation in string theory,
or string theory in de Sitter space.
If we assume the existence of an effective
action in these environments, though,
we can compute by evaluating terms in the effective
action from the string theory S matrix in
flat space ($H=0$).

These models have four real supercharges and hence have
no constraints on the kinetic term in the
four dimensional effective action.  On the other hand,
sixteen or more supercharges  would require no
renormalization of the kinetic term.   So as
$m_{CY} \rightarrow 0$ and flat ten dimensional space
is recovered, the higher derivative terms in the
effective action must vanish.  So we expect effects
in the propagator \corrat\ of size $m_{CY}^2 H^2/m_s^4$.
For $m_{CY}\sim m_s$  this becomes
$H^2/m_s^2 \sim 10^{-11}$.  This is unobservable.

To find effects closer to the threshold of detectability
we must enter the realm of strongly coupled
string theory, where the fundamental mass scale
can be much less than $m_4$.

\def\tphi{\tilde{\phi}}


\lref\cvetic{M. Cvetic, G. Shiu, and A. Uranga,
``Three Family Supersymmetric Standard--Like Models from Intersecting
Brane Worlds,'' Phys. Rev. Lett.  {\bf 87}, (2001) 201801,
hep-th/0107143; and M. Cvetic, G. Shiu, and A. Uranga,
``Chiral Four-Dimensional N = 1 Supersymmetric Type IIA Orientifolds from
Intersecting D6-Branes,''
Nucl. Phys. B {\bf 615},(2001) 3,
hep-th/0107166.}



\lref\aadd{I. Antoniadis, N. Arkani-Hamed, S. Dimopoulos
and G. Dvali, ``New Dimensions at a Millimeter to a Fermi
and Superstrings at a TeV,'' \pl\ {\bf B436}\ (1998) 257;\
hep-ph/9804398.}

\newsec{Large effects in string and M-theory}

We have shown that
new physics at a scale $M$ leads to
the following expression for quantum
fluctuations of the inflaton:
\eqn\formofcorr{
    \langle \delta\phi^2 \rangle = \frac{H^2}{4\pi^2}
    \left(1 + \CX \frac{H^2}{M^2} +
    \cdots \right)\ .
}
The second term in brackets is the leading correction
to the standard, free-field expression used in inflationary cosmology.
$\CX$ is a model-dependent, dimensionless number related
to the coefficients in the effective action \effact.
It may get contributions
from phase space factors in loop integrals,
sums over heavy particles coupling to the
inflaton, and so on.

As we will argue in the next section, this
correction is potentially observable as a correction
to a well-known consistency condition
on the tensor and scalar fluctuations of the CMBR.
We believe such an effect is measurable in principle if
\eqn\measurability{
    \CX \frac{H^2}{M^2} \sim 0.1 - 1\ .
}
It is hard to be more precise with this number, as
it depends on the measurability of the B-mode polarization,
which is not yet well understood.

The Hubble constant $H^2$ can be calculated, or hopefully measured
in polarization experiments. The current upper bound
from COBE, current degree-scale anisotropy experiments,
and large-scale structure data is \hbox{$H = 7\times 10^{13}$\
GeV} \kamkos. In 4D GUT models, $M=m_4$, $\CX \ll 1$, and the
correction is unobservable.\foot{Models for which the particle
coupling to the inflaton becomes light some $\phi = \phi_0$ during
the inflationary epoch \kolbresonant\ lead to an observable effect
at a particular angular scale on the sky; this will be
observationally distinct from the effects we discuss here.}
However, in most phenomenologically viable string and M-theory
models, the fundamental scale $M_f$ -- either the
higher-dimensional Planck scale or the string scale -- is lower
than the $4D$ Planck scale $m_4$ by up to two orders of magnitude
\refs{\dsscales,\kscales,\witstrong}. If we compactify a
$d$-dimensional theory with Planck scale $M_f$ on a
$(d-4)$-dimensional manifold $X_{d-4}$ with volume $V_{d-4}$, then
$m_4^2 = M_f^{d-2} V_{d-4}$. The high scale $m_4$ is not a
dynamical scale, but rather an artifact of the large volume of the
compactification manifold.

In these models we might expect $M=M_f$.
However, so long as the Hubble scale
is lower than the compactification
scale, $4D$ effective field theory
still applies.  The effect on \formofcorr\ of
integrating out a given four-dimensional
field still gives
$M=m_4, \CX < 1$.

But higher-dimensional models
have several new features which
can significantly enhance the corrections
to \formofcorr.  First, the corrections
in \formofcorr\ arise from nonrenormalizable
gravitational couplings which become
large at high energies.  Thus high-scale
physics--in particular the large numbers of
particles above the Kaluza-Klein threshold--contributes
significantly in loops.
Secondly, the existence of tensor fields in 10-
and 11-dimensional models leads to a large factor $\CX$ from
summing over polarizations of these fields.

In almost all of the models we are interested
in, the dominant effects arise from supergravity modes.
The loop integrals appear highly divergent;
but for the effects we are calculating
they are cut off by
either the restoration of maximal
supersymmetry (to 16 or 32 unbroken supercharges),
or by the soft ultraviolet behavior
of the fundamental theory. The result is highly
model-dependent, and the numbers we arrive at
by no means constitute a precise prediction.
Nonetheless we can estimate whether
the correction in \formofcorr\ is observable.
To that end, we will begin this
section by estimating $\CX H^2/M^2$
as a function of the compactification
radii and the cutoff.  We will then
analyze a variety of supersymmetric $N=1, d=4$
models in string and M-theory and estimate
the size of the one-loop contribution to
\formofcorr\ in each.  Readers who
are less theoretically inclined (or simply
impatient) will find the results of this section
in the following paragraphs above \S4.1;
they may then skip to \S5\ where the experimental
consequences are discussed.

\subsubsection{Summary of results}

Our strategy will be as follows.
The first two models we analyze --
M-theory on $X_7 = X_6 \times S^1/\IZ_2$ \horwit,
also known as  \Hor-Witten theory,
and M-theory on a manifold of $G_2$ holonomy --
can be made consistent with the unification prediction
of \unify, by keeping all scales including the eleven-dimensional
Planck scale to within an order of magnitude of the
unification scale, $\mg = 2 \times 10^{16}$~GeV.
In this case we
will find that $\CX H^2/M^2 \sim 10^{-7}$, which
is unobservably small.

However, if we give up perturbative
unification, we can increase the volume of the compactification
manifold and decrease the fundamental Planck scale.
We examine such models under three constraints.
First, the inflationary dynamics must remain four-dimensional.
This puts an upper limit on the size of the
compactification manifold, on the order of $1/H$.
Secondly, if we assume that the
energy density during inflation is
constant in $d$ dimensions,
then it must be lower than $m_d^d$, where $m_d$ is the
$d$-dimensional (reduced) Planck scale.  We will find that this
also places an upper limit on the compactification volume.

Finally, the four-dimensional gauge coupling must remain $\alpha
\sim 1/25$ in order that the standard model couplings are roughly
correct at a $TeV$.  The origin of gauge dynamics in a given
model, combined with the constraint on $\alpha$, affects how many
dimensions can be made large.  In M- and F-theory models, gauge
dynamics arises on singularities or on branes, both at finite
codimension in the compactification manifold. If the singularity or brane
lies on a $k$-dimensional submanifold $\Sigma \subset X$, $\alpha
= V_\Sigma M_f^k$ and $V_\Sigma$ is fixed.  The number of
dimensions which may be made large is then the codimension $d-k$
of the brane or singularity, so the models with the greatest
chance of giving rise to observable corrections in \formofcorr\
are those with the highest $d-k$.

In M-theory on $X_6 \times S^1/\IZ_2$ the
gauge dynamics occurs on the boundaries of
$S^1/\IZ_2$ which have codimension one.
The volume of $X_6$
is constrained, and we cannot decrease the
size of the interval $S^1/\IZ_2$ low enough to
make the correction term in \formofcorr\
observable.  Manifolds with
$G_2$ holonomy are in much better shape.
The gauge dynamics lies on singularities of codimension
four \gtwochiral.
We can increase the volume of the transverse
manifold such that the eleven-dimensional
Planck scale is $m_{11} \sim H$.

We then move to ten-dimensional
type I models. In the simplest
such models the gauge degrees
of freedom propagate in ten dimensions.
The compactification manifold can be made
large consistently with $\alpha = 1/25$ by
adjusting the string coupling.
But this coupling is weak, so that ten-dimensional
physics is controlled by very soft string physics
and the correction in \formofcorr\ is unobservable.

One may also study models for which the
gauge degrees of freedom propagate on branes.
Two such models consistent with
$\CN=1$ supersymmetry in four dimensions
are \Hor-Witten models with the gauge dynamics
arising on M5-branes wrapped on
Riemann surfaces, and F-theory models
with the gauge dynamics arising on
D3-branes.  For both of these models,
the strongest constraint is that imposed by
sub-Planckian energy densities.
Up to the model-dependent factor $\CX$,
the constraints on the corrections in
\formofcorr\ lead to estimates
for $\CX H^2/M^2$ that are within a factor of a few of
the estimate for
manifolds of $G_2$ holonomy, so we
will not discuss these other models further.

All of these estimates are model-dependent and imprecise.
In particular, we will argue below that
the loop expansion is starting to break down
as the corrections in \formofcorr\ start to become observable.
It is easy to imagine these effects
changing our estimates
by an order of magnitude in
more precise calculations.  For this reason we
parameterize our results in terms of $H^2/M^2$
and $\CX$ separately.

\subsec{Notation}

First we specify our notation: the dimensionful
gravitational coupling $\kappa_d$ is the coefficient
of the Einstein term:
\eqn\elevenein{
    \CL = \frac{1}{2\kappa_d^2} \int_{X_d} d^d x \sqg R\ .
}
We define two versions of the
Planck mass (differing by a numerical factor):
\eqn\usuallength{
    2\kappa_d^2 = (2\pi)^8 \ell_d^9\ = \tilde{\ell}_d^9 ;\ \ \ \ \
    M_d = 1/\ell_d\ , m_d = 1/\tilde{\ell}_d\ .
}
When we compactify on a manifold $X_{d-4}$
with volume $V_{d-4}$, the four-dimensional Planck scale is:
\eqn\fourdplanckind{
    m_4^2 = 2 m_d^{d-2} V_{d-4}\ .
}

In ten-dimensional string theories the gravitational
coupling can be written via the string scale as:
\eqn\stringscaledef{
    2\kappa_{10}^2 = \gs^2 \ap^4 = \frac{\gs^2}{\ms^8} =
    \gs^2 (2\pi)^7 \ell_s^8
    = \gs^2 \frac{(2\pi)^7}{M_s^8}\ .
}
The string tension is $T = 1/2\pi\ap$ and a string oscillator mode
carries energy $1/\sqrt{\ap}$.

\subsec{Corrections to the propagator in
higher-dimensional theories}

Ideally we could
choose a string model and simply calculate the
one-loop corrections to \formofcorr\ in perturbative string theory.
However, string theory in
approximately de Sitter backgrounds
is poorly understood.  Furthermore,
we will find that the effects
of high scale physics
are closest to observability
in M- and F-theory models.

However, supergravity remains a good approximation
in the calculations we are interested in.  We
will begin by simply studying a scalar field
coupled to a $d$-dimensional graviton
on a $(d-4)$-dimensional torus.
This may seem nonsensical as
the loop integrals will be badly divergent.
However, the corrections to the
$p^2$ and $p^4$ terms in the
propagator vanish in supersymmetric theories
when 16 or 32 supercharges are unbroken,
at energies above the compactification
scale.  Therefore, supersymmetry
cuts off the otherwise highly divergent
amplitudes without our needing
to appeal to the ultraviolet
physics of M- or F-theory.  The scale of
the cutoff will be set by the
scale at which the full
supersymmetry of the underlying theory is restored.\foot{Of 
course supersymmetry is also broken by the vacuum
energy.  However it is restored for momenta
$k > H$.  The corrections we will discuss will
arise from momenta much larger than $H$.}

Because the loop integrals are dominated by
energies near the cutoff, well above the
compactification scale,
we can ignore the effects of the curvature and
topology of $X_d$.  We will therefore
estimate the correction in \formofcorr\
by coupling the inflaton to the d-dimensional
graviton on a rectangular $(d-4)$-dimensional
torus.  For our purposes,
the effects of the actual geometry
can be summarized in terms
of the model-dependence of $\CX$ in \formofcorr.

The Lagrangian for a massive scalar minimally
coupled to the d-dimensional graviton is
\eqn\inflatonaction{
    S = \frac{1}{2} \int d^{d} x \sqg \left( g^{a b}
    \p_a \phi \p_b \phi + m^2\phi^2\right)\ .
}
The metric $g$ can be written in terms of
the background metric $\eta$ (which we
take to be flat) and a small fluctuation:
\eqn\metexp{
    g_{ab} = \eta_{ab} + h_{ab}\ .
}
$S$ can be expanded in powers of $h$ using the formulae
\eqn\varformet{
\eqalign{
    \delta \sqg &= \half\sqg g^{ab}\delta g_{ab}\cr
    \delta g^{ab} &= -g^{ac}g^{bd}\delta g_{bd}\ .
}}
This will lead to nonrenormalizable couplings of the
form $h(\p\phi)^2$ and $h^2 (\p\phi)^2$:
\eqn\intact{
\eqalign{
    \delta S = \half \int d^d x\left(\right.
        &T_3^{ab,mn}h_{ab}\p_m\phi\p_n\phi
        + T_4^{abcd,mn}h_{ab}h_{cd}\p_m\phi\p_n\phi
        \cr
    &\left.
    \ \ \ \ \  + M_3^{ab}h_{ab}\phi^2 +
    M_4^{abcd}h_{ab}h_{cd}\phi^2
    \right)\ ,
}}
where
\eqn\couplingtensors{
\eqalign{
    T_3^{ab,mn} & = \half\eta^{ab}\eta^{mn}
        - \eta^{ma}\eta^{nb}\cr
    T_4^{abcd,mn} &= -\half \eta^{ab}\eta^{mc}\eta^{nd}
    + \frac{1}{8}\eta^{ab}\eta^{cd}\eta^{mn}
    - \frac{1}{4}\eta^{ac}\eta^{bd}\eta^{mn}
    + \half\eta^{mc}\eta^{ad}\eta^{nb}
    + \half\eta^{ma}\eta^{nc}\eta^{bd}\cr
    M_3^{ab} &= \half\ m^2 \eta^{ab}\cr
    M_4^{abcd} &= \frac{1}{8} m^2 \eta^{ab}\eta^{cd}
        - \frac{1}{4}\ m^2\eta^{ac}\eta^{bd}\ .
}}

The propagator for $h$ is, in de Donder gauge:
\eqn\gravitonprop{
    \langle h_{ab} h_{cd} \rangle
    = \frac{1}{m_4^2 k^2}
    \left( \eta_{ac} \eta_{bd} + \eta_{ad}\eta_{bc}
        - \frac{2}{d-2}\eta_{ab} \eta_{cd} \right).
}
The two one-loop diagrams are shown in Fig. 2.
The diagram on the left will contribute both wavefunction renormalization
terms and $p^4/\Lambda^2$ corrections to the propagator, while the
right-hand diagram will give further wavefunction renormalization
corrections.

\ifig\feynman{Two one-loop diagrams important for the computation.}
{\epsfxsize2.0in\epsfbox{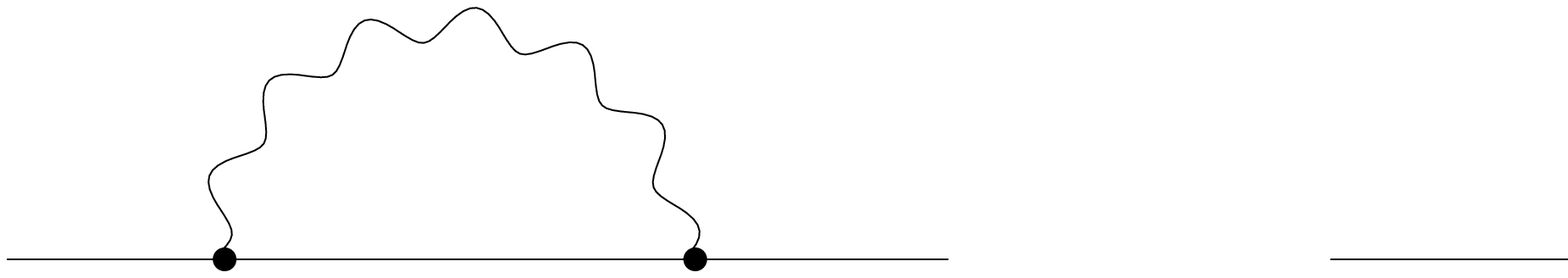}}

We are interested in the divergent part of the loops
with loop momenta of order $k \gg H$.  Therefore,
we can approximate the
$H$-dependence of the propagators at tree level via the
first two terms in \effact, which amounts to shifting
all of the masses by $m^2 \to m^2 - 2H^2$.  At the end
we will take the leading correction in $H^2/\Lambda^2$.

The left-hand diagram leads to the correction:
\eqn\propcorrone{
    D_1(p) =
        \frac{1}{m_4^2} \sum_{n}
        \int \frac{d^4 k}{(2\pi)^4}
        \frac{p^2 (p-k_n)^2 + p\cdot(p-k_n) m^2 + m^4}
        {\left(k_n^2 + m^2 - 2H^2\right)
        \left((p-k_n)^2 - 2H^2\right)}
}
and the right-hand diagram to the correction:
\eqn\propcorrtwo{
    D_2(p) = -\frac{1}{4}(d^2 + d - 8)
    \frac{1}{m_4^2} \sum_n \int \frac{d^4 k}{(2\pi)^4}
    \frac{(p^2 + m^2)}{k_n^2 - 2H^2}\ .
}
The sum over $n$ is over
Kaluza-Klein momenta, and $k_n$ denotes the
full eleven-dimensional momentum
of the internal graviton propagator.

The four-dimensional integrals in \propcorrone,\propcorrtwo\
are already quadratically divergent, and
the Kaluza-Klein sum only increases
the degree of divergence.  So long as the
cutoff is more than a few
times the Kaluza-Klein scale,
we can approximate this sum by an integral:
\eqn\kkfactor{
    \sum_{n} = \frac{V_{d-4}}{(2\pi)^{d-4}} \int d^{d-4} k\ ,
}
where $V_{d-4}$ is the radius of the $T^{d-4}$.
The eleven-dimensional momentum integrals
are highly divergent and
dominated by the UV end of the integral,
near the cutoff $\Lambda \gg H$.  We can
therefore expand the integrand in powers of $H^2/k^2$.
After subtracting the $H$-independent wavefunction renormalization
correction, the most divergent terms in this expansion are:
\eqn\fulldpropcorr{
    D(p) = -\frac{n_d V_{d-4} H^2 (p^2 + m^2)}{m_4^2}
    \int \frac{d^d k}{(2\pi)^d} \frac{1}{k^4}
    + \frac{ V_{d-4}p^4}{m_4^2} \int
        \frac{d^d k}{(2\pi)^d} \frac{1}{k^4}\ .
}
The first term leads to an $H$-dependent
wavefunction renormalization, and the
second to a $p^4$ term in the propagator.
The factor $n_d$ arises from the sum over graviton
polarizations.  Since de Donder gauge is not
complete, we must subtract off the ghosts.
The result should be the number of physical
graviton polarizations, $(d-1)(d-2)/2 - 1$.
In theories with 32 unbroken supercharges
before compactification, the graviton
supermultiplet contains additional scalars,
gauge fields, and tensors, so that when
we include all of the bosonic degrees of freedom
we will find $n_d = 128$.

The size of \fulldpropcorr\ depends strongly on the cutoff.
Na\"{\i}vely one expects this cutoff to be $M_f$.
If $M_f = m_d$, then
\eqn\ddimred{
    V_{d-4}m_d^{d-4} = \frac{m_4^2}{M_d^2}
}
and so we can write $M=m_d$ in \formofcorr.
However, the
relation between $\Lambda$ and $M_f$
is model-dependent, and may involve factors
of $2\pi$ and other dimensionless numbers.
Because these factors are raised to high powers,
they can have a significant
effect on the size of the correction in \formofcorr.
For now we will set
\eqn\fundtocut{
    \Lambda = c m_d\ ,
}
with $c \sim \CO(1)$ parameterizing the
model-dependence.

After performing the momentum integrals in \fulldpropcorr,
we find:
\eqn\newcorrect{
    D(p) = - \frac{2 n_d \pi^{d/2} c^{d-4}}
    {(d-4)(2\pi)^d
    \Gamma\left(\frac{d}{2}\right)}
        \frac{H^2}{M_f^2} \left( p^2 + m^2 \right)
    + \frac{ 2 \pi^{d/2} c^{d-4}}{
    (d-4)(2\pi)^d \Gamma\left(\frac{d}{2}\right)}
    \frac{p^4}{m_d^2}\ .
}
Since $n_d \sim 100$, the wavefunction renormalization
term will dominate, and we will find that
the coefficient $\CX$ in \formofcorr\ will take the value:
\eqn\normalizecoeff{
    \CX = \frac{2 n_d \pi^{d/2} c^{d-4}}
    {(d-4)(2\pi)^d
    \Gamma\left(\frac{d}{2}\right)} ,
}
while $M = M_f$.

These estimates are hardly precise.
In addition to the model-dependence we have discussed,
the loop expansion will begin to break down in models
where the corrections in \formofcorr\ are
close to observability.  For these models $c\geq 1$
in \fundtocut.
If the fundamental scale is $M_f = m_d$,
where $m_d$ is the d-dimensional Planck scale,
then the dimensionless gravitational coupling governing
loop corrections will be:
\eqn\gravcoupling{
    g_{grav}^2 =
        \left( \frac{\Lambda}{m_d} \right)^{d-2}\ .
}
Once $\Lambda \sim m_d$, $g_{grav}^2 \sim 1$.
Nonetheless we will assume that the one-loop
answer gives a rough estimate of the size of
the corrections in \formofcorr.

We will still try to be careful
about numerical factors.
This may seem perverse given the above discussion.
However, these factors are often
raised to high powers, so that they contribute
appreciably to our order-of-magnitude
estimates.

The remainder of this section will amount to
estimates of the magnitude of \formofcorr\ in a variety
of string and M-theory models, with these caveats
firmly in mind.

\subsec{Physical constraints on compactifications}

In the minimal supersymmetric
standard model, the running
strong, weak and electromagnetic couplings unify at
\eqn\unifpred{
    \ag = \frac{g^2}{4\pi} \sim \frac{1}{25}
}
at a scale of order
$\mg\sim 2\times 10^{16}\ {\rm GeV}$ \unify.
This is strong evidence for grand unification
at that scale.  Nonetheless it still
indicates a small hierarchy between
$\mg$ and $m_4$.

In traditional string phenomenology, one
starts with ten-dimensional type I
or heterotic string theories, which
have 16 unbroken supercharges.  One then
chooses a six-dimensional Calabi-Yau
manifold $X$ with volume $V_X \sim \mg^{-6}$,
which preserves $N=1$ SUSY at the compactification scale.
$\mg$,\ $\ag$ and $m_4$ are
computable functions of $V_X$, the string scale $\ms$
and the string coupling $\gs$, and one may adjust the
compactification parameters in order to
match the unification predictions of \unify.

For type I models, the measured values
of $\ag$,$\mg$, and $m_4$ can be achieved in models
with weak string coupling.
For heterotic models, the observed couplings and
scales are incompatible with
weak string coupling \witstrong.
One may try to work at strong heterotic
coupling, but it is not clear
that the expressions for the gauge couplings
are correct.

Instead we can appeal to string duality \witstrong.
The strong coupling limit of the $SO(32)$ string
is weakly coupled type I string theory \hettypeI.
The strong coupling limit of the
$E_8\times E_8$ heterotic
string compactified on $X$
is M-theory on $X\times S^1/Z_2$ \horwit.
In this latter limit, gauge coupling unification is
compatible with a background
well described by 11-dimensional
supergravity \witstrong.

We will also study M-theory
on a manifold of $G_2$ holonomy,
and weakly coupled
type I string models.
We will find that \Hor-Witten theory
(with the standard model
as a subgroup of $E_8 \times E_8$)
and weakly coupled type I
models do not give rise to observable corrections
in \formofcorr, in any reasonable regimes
of parameter space.  It appears that
for such corrections to be observable, the
dynamics must be strongly coupled and the
standard model should live on a brane
or singularity with high codimension.

In the remainder of this section we
will discuss a variety of models
which have low-energy gauge dynamics and
a fundamental scale lower than $m_4$,
and estimate the size of corrections
to \formofcorr.
We will spend the most time on
\Hor-Witten models.  We will
then discuss M theory on manifolds
of $G_2$ holonomy and \Hor-Witten theory
type $I$ models.

In models consistent with coupling unification,
the correction to \formofcorr\ will
turn out to be too small to be observed.
We will therefore examine a wider class of models
under the following constraints.  First,
the four-dimensional Planck scale
must be that given by experiment.
Secondly, although we have
given up coupling unification,
the gauge coupling at the
fundamental scale must be on the order of
$\alpha = 1/25$, to get roughly the correct
standard model couplings at a $TeV$.  Thirdly,
we will demand that inflationary dynamics
be truly four-dimensional.  The upper limit on the
compactification volume is set by demanding that the
Kaluza-Klein momenta be larger than
the deSitter temperature, $T_{dS} = H/2\pi$,
so that the dynamics
of quantum inflaton fluctuations
remains four-dimensional.
If we imagine compactification on a circle with circumference
$L$,  this condition means that $L < (2\pi)^2/H$.
For a manifold $X_k$ with volume $V_k$, we take this
to mean that
\eqn\volumeupperlimit{
    V_k \leq \frac{(2\pi)^{2k}}{H^k}\ .
}

Finally, we demand that the $d$-dimensional energy density be
sub-Planckian.  Let us assume that the energy
density responsible for inflation is constant over
the compactification manifold $X_{d-4}$.
Denoting the $k$-dimensional energy density by
$E_{(k)}^k$,
\eqn\energydenrel{
    E_{(4)}^4 = 3 H^2 m_4^2 = E_{(d)}^d V_{d-4} \ ,
}
which implies
\eqn\energyratio{
    \left( \frac{H}{m_d}\right)^2 = \frac{1}{3}
    \left( \frac{E_{(d)}}{m_d} \right)^d\ .
}
Therefore we demand\foot{
We ignore the factor of $1/3$; it disappears if we
allow e.g. $E_{(d)} = 1.2 m_d$ which we cannot rule out at this
crude level.}
\eqn\maximumratio{
    \left( \frac{H}{m_d}\right)^2 \leq 1\ .
}
Since $m_4$ is fixed, Eq. \fourdplanckind\ ties a lower limit
on $m_d$ to an upper limit on $V_{d-4}$.
Depending on the model at hand,
this bound may be more or less stringent
than \volumeupperlimit.

In our study of perturbative type I models,
we will also demand that the energy density
$E_{(10)}^{10} \leq \ms^{10}$.  At higher energy densities
stringy physics is not understood.

We will find that for models which give
measureable correction terms in \formofcorr,
the fundamental scale
$m_d \sim H \sim 7 \times 10^{13}$~GeV.
With such a low scale we have to worry again about
proton decay.  In GUT models dimension-six operators
suppressed by $1/\mg^2$
lead to proton lifetimes close to the experimental
lower bound, close enough to model-dependent factors
to rule out models.  Dimension-six operators
suppressed by $1/H^2$ will lead to proton decay which is
10 or 11 orders of magnitude more rapid than
if they were suppressed by $1/\mg^2$.  If one
is able to forbid operators below dimension seven,
then higher-dimensional operators
suppressed by powers of $1/H$ will lead
to phenomenologically acceptable lifetimes.
One could achieve this, for example, if
some discrete subgroup of the $U(1)$ baryon
number symmetry was gauged,
along the lines of \refs{\add, \aadd}.
Since we are not studying our models in detail
we will leave this issue aside.

\subsec{The \Hor-Witten model}

Compactifications of M-theory on $X_{11} = X_{10}\times S^1/Z_2$
were the first known M-theory models
with chiral gauge dynamics \horwit.  These
models can be described relatively explicitly,
so we will spend the greatest amount of time on them.
In addition, in models which realize $N=1$ supersymmetry,
the explicit pattern of supersymmetry breaking
means that moduli of the compactification manifold
are good inflaton candidates \refs{\tomcosm,\tomcosmrev}
as we will review below.

If $X_{10} = \IR^4 \times X_6$ and $X_6$ is Calabi-Yau,
the theory has four unbroken supercharges in four dimensions.
One $E_8$ gauge multiplet is localized on each end of
the interval.  The gauge couplings are:
\eqn\tengauge{
    \sum_{i=1}^2 \frac{1}{8\pi
    \left(4\pi\k11^2\right)^{2/3}}
    \int_{M_{10,i}} \sqg F_i^2
}
where the sum is over the two boundary components.
Upon compactification on $X$, anomaly cancellation
will require gauge field configurations which
break this gauge group further; generally
one breaks one of the $E_8$ groups
to the GUT group and then to the standard model
gauge group, while the other $E_8$ is the
gauge symmetry of a hidden sector.

Without going into great detail, we
can see that these models can
match the predicted coupling unification
in a regime where all scales, including the
fundamental scale, are close to $\mg$ and
supergravity is valid.

The GUT group is broken
to the standard model gauge group
by visible sector gauge field configurations on $X$ --
\cf\ \gsw\ for a discussion.  Therefore
we let  $L_{CY} = V_X^{1/6} = \mg^{-1}$.
Newton's constant $G_N$ and the GUT coupling $\ag$
can be written as \witstrong:
\eqn\fourdcouple{
\eqalign{
    \frac{1}{8\pi G_N} &= \frac{V_X L_{11}}{\k11^2}\cr
    \ag &= \frac{g_{GUT}^2}{4\pi}
        = \frac{\left(4\pi\k11^2\right)^{2/3}}{2V_X}
}}
With the above values of $\ag$ and $\mg$,
we find:
\eqn\elevendscales{
\eqalign{
    m_{11} & \sim 2 \mg\cr
    \m11 & \sim 10 \mg\cr
    \frac{1}{L_{11}} & \sim 0.01 \mg\ .
}}
Therefore although the heterotic
coupling is strong, this compactification is well
described by $11d$ supergravity.
Note that we do not have to postulate large hierarchies
between the GUT and fundamental scales.  The
largest hierarchy is between $\m11$ and $1/L_{11}$.
If we take the ratio between
$\m11$ and the mass gap of the Kaluza-Klein excitations
with momentum along $S^1/\IZ_2$:
\eqn\betterratio{
    m_{11}/m_{KK} = m_{11} L_{11} /\pi \sim 60\ ,
}
so the Kaluza-Klein scale is about an order of magnitude off
from the GUT scale.

The expansion parameter in these models is
$(2\k11^2)^{2/3}/V_X$.  The assumption that
the geometry is a simple product $X\times S^1/\IZ_2$
holds only at lowest order.
To next order in our expansion parameter the product is
warped; $V_X$ depends on the coordinate
$x_{11}$ along $S^1/\IZ_2$ \witstrong.
A natural size for $L_{11}$ is that for which
the volume vanishes at the end of the interval
where the hidden sector gauge group resides.
One can then imagine strong coupling effects
leading to supersymmetry breaking and the
stabilization of moduli \refs{\witstrong,\gluino}.
$L_{11}$ determined this way depends on the topology of the
$E_8 \times E_8$
gauge field configurations on $X$, and
on the K\"ahler moduli of $X$.
For reasonable choices of both, $L_{11}$
is consistent with \elevendscales.

In this model maximal supersymmetry is broken to
$N=1$ in ten dimensions at the fixed points of $S^1/\IZ_2$,
and then to $N=1$ in $d=4$ by the
compactification on $X_6$.  The
cutoff in \kkfactor\ should be roughly $V_X^{-1/6}$, so
we might expect the cutoff to be on the order
of $1/V_X^{1/6} \sim \mg$.  Again, the precise value of
$\Lambda$ is highly model-dependent.  In a sufficiently
anisotropic Calabi-Yau we can raise this scale.
We will take it to be the fundamental UV cutoff that quantum-mechanical
M-theory is expected to provide.

This cutoff can be estimated by studying four-graviton
scattering at one loop \ruts; since the amplitude is protected by
supersymmetry, it can be calculated in string theory
and extrapolated to strong coupling.  The computation
is cutoff dependent in supergravity.
If we define the cutoff $\Lambda_{11}$ by matching the
supergravity result to the
finite M-theory result, then \ruts:
\eqn\cutoff{
    \Lambda_{11} = 2^{4/9}\pi^{11/9} m_{11} \sim 5 m_{11}\ .
}

\subsubsection{Inflaton dynamics in the \Hor-Witten model}

In the \Hor-Witten models, the moduli of $X_6$
are natural inflaton candidates.
A simple argument due to Banks
\refs{\tomcosm,\tomcosmrev}\foot{With
many caveats, extensively discussed in
\tomcosmrev.}
shows that the pattern of supersymmetry
breaking in \Hor-Witten models can lead
to an inflaton potential with the right properties.
M-theory compactified on a Calabi-Yau threefold
$X$ has eight supercharges, and the moduli
of $X$ are exactly flat directions, protected
by supersymmetry.  Upon further compactification
on $S^1/\IZ_2$, supersymmetry is broken
to $N=1$ in four dimensions {\it at the boundaries
of the interval}.  Superpotentials for the moduli of
$X$ can arise only on the boundaries.

Let $\tphi^A$ be the (complex) moduli of $X$ in M-theory,
describing sizes of various cycles
of $X$ in units of $\m11$.
In four dimensions the kinetic term is:
\eqn\modulilagr{
    \CL_{kin} = \frac{1}{2\k11^2}V_X L_{11}
    \int d^4 x G_{A\bar{B}}(\tphi)\p\tphi^A
        \p\bar{\tphi}^{\bar{B}}\ .
}
where $G_{AB}$ is the dimensionless metric
on the moduli space of $X$.
The factor in front of the integral also
multiplies the 4-dimensional Einstein term, which is expected since the
moduli are simply components of the metric in
$X$. This is just the ``reduced'' $4d$ Planck mass $m_4$.

The canonically normalized scalar fields in four dimensions
are:
\eqn\canonical{
    \phi^A = m_4\tphi^A\ .
}
In $N=1$ language, $G$ is the derivative of the
K\"ahler potential
\eqn\metricdef{
    G_{A \bar{B}} = \p_A\p_{\bar{B}} K
}
where the derivatives are with respect to the canonically
normalized fields.  The fact that $G$ is dimensionless
and of order one means that we can write $K$ in terms
of a dimensionless order one potential $\tilde{K}$:
\eqn\redk{ K = m_4^2 \tilde{K} }
so that
\eqn\metricagain{
    G_{A\bar{B}} = \tilde{\p}_A\tilde{\p}_B \tilde{K}
}
where $\tilde{\p}$ is a the derivative with respect to $\tphi$.

$N=2$ SUSY is broken to $N=1$ by the boundaries
of $S^1/\IZ_2$.  Fundamental physics
on these boundaries is still controlled by $m_{11}$,
so that the superpotential will have the form:
\eqn\modulisuper{
    \CL_{super} = m_{11}^3 \int d^2\theta d^4 x
        w(\tphi) + {\rm h.c.}
}
The bosonic potential in $N=1$ supergravity
arising from this superpotential is:
\eqn\sugrapotent{
    V(\phi) = e^{\tilde{K}}\frac{m_{11}^6}{m_4^2}\left(
    G^{A\bar{B}}\tilde{D}_A w \tilde{D}_{\bar{B}}
    \bar{w} - 3 |w|^2 \right)
    = \frac{m_{11}^6}{m_4^2} {\cal V} \left(\frac{\phi}{m_4}\right)
    \equiv M^4 {\cal V} ,
}
where
$$ \tilde{D}_A w = \tilde{\p}_A w + \tilde{\p}_A \tilde{K} w\ . $$

For a successful model of inflation,
$\phi$ must roll slowly for approximately $60$
e-foldings, and the fluctuations in $\phi$
must generate the density perturbations
measured by COBE,
$\delta\rho/\rho \sim 5\times 10^{-5}$.
We can use this requirement to compute $M$ \modular.
We will choose our coordinates so that a single
coordinate $\tphi$ parameterizes the trajectory
in the moduli space travelled during the inflationary
epoch.

We rewrite \efolds\ in the present context:
\eqn\genefoldings{
    N_e = \frac{1}{2m_4^2} \int_{\phi_e}^\phi
        \frac{V}{\p_\phi V} \sim 60\ ,
}
using the slow-roll expression
$$ H^2 = \frac{V}{3m_4^2} \ . $$
Here $\phi_e$ is the vev of the inflaton at the end
of inflation, and $\phi$ the vev 60 efoldings
prior to that.
Assuming $V$ does not change much during inflation
we can approximate \genefoldings\ by:
\eqn\approxefoldings{
    \frac{\phi - \phi_e}{m_4^2} \frac{{\cal V}}{\partial_\phi {\cal V}}
        \sim 60\ .
}
If we let $(\phi-\phi_e) \sim m_4$
and solve for ${\cal V}/\partial_\phi {\cal V}$,
we can use \dencontf\ to solve for $M$ and $\m11$:
\eqn\energyscale{
\eqalign{
    M & \sim 3 \times 10^{-3}\ m_4 =
        7 \times 10^{15} {\rm GeV} \cr
    m_{11} &\sim 5 \times 10^{16} {\rm GeV} \ .
}}
$M$ is close to the
unification scale $\mg$, and the value of
$m_{11}$ predicted here is close to that in
\elevendscales.  Within our crude
set of approximations we can take this
as an estimate of $m_{11}$ independent
of \elevendscales.

Since these numbers are rough estimates,
we will use the experimentally determined
upper bound $H$ in our estimates of
$H^2/M^2$.

\subsubsection{Corrections to the inflaton propagator}

Eq. \cutoff\ implies that
$c = 2^{4/9}\pi^{11/9}$ in \normalizecoeff.
The correction in \formofcorr\ is determined by: (here $d = 11$, $n_b = 128$)
\eqn\hwfoc{
    M^2 \sim m_{11}^2\ ,\ \ \ \ \ \CX \sim 0.1\ .
}
Using the experimental upper bound on $H$,
and $m_{11}$ as given in \elevendscales,
\eqn\hwunifno{
    \frac{H^2}{M^2} \sim 10^{-6}
}

While this is better than the
result expected from four-dimensional GUT models,
it is not close to observable.  If we were willing
to give up unification at $\mg$ and require only
that the gauge couplings satisfy $\alpha\sim 1/25$ and
that the inflationary dynamics be four-dimensional,
we can have a smaller value of $m_{11}$ and the corrections
in \formofcorr\ will be larger.  (If we push these constraints
to their limits, the eleven-dimensional energy density
is still sub-Planckian).  Note that for such models
the arguments in \refs{\tomcosm,\tomcosmrev}
will cease to generate
inflaton potentials with $\CM\sim \mg$,
as we must push $m_{11} < \mg$ for
corrections in \formofcorr\ to be observable.
We will have to assume that
such potentials are generated by
four-dimensional gauge theory effects.

The correct four-dimensional Planck scale,
\eqn\widermp{
    m_4^2 = \alpha^{-1} \left(4\pi\right)^{2/3}
        L_{11} m_{11}^3\ ,
}
The constraint that
the inflaton fluctuations remain four-dimensional is:
\eqn\largeint{
    L_{11} < \frac{1}{\gamma H};\ \ \ \ \ \
    \gamma > \frac{1}{(2\pi)^2}\ ,
}
while $\alpha \sim 1/25$ constrains the
Calabi-Yau volume.  Then
\eqn\hwnewratio{
    \frac{H^2}{m_{11}^2} =
    \frac{(25)^{2/3}(4\pi)^{4/9}}{\gamma^{2/3}}
    \left(\frac{H}{m_4}\right)^{4/3}\ ,
}
Assuming also $\gamma = 1/(2\pi)^2$ and $H = 7 \times 10^{13}$~GeV,
we find that:
\eqn\hwlowplanck{
    m_{11} \sim 6 \times 10^{15} {\rm GeV}\ ,
    \ \ \ \ \ 1/L_{11} \sim 10^{12}\ {\rm GeV}\ ,
}
so that
\eqn\hwnewcorr{
    \frac{H^2}{M^2} \sim 10^{-4}\ ,\ \ \ \ \ \CX \sim 0.1\ .
}
This is a considerable improvement, but it is still
unobservable.  We will find below that if the
gauge dynamics are restricted to a lower-dimensional
brane, more directions transverse to the brane
may be made large, and the fundamental scale
can be lowered further still,
while keeping the four-dimensional Planck scale fixed.

\subsec{$G_2$ manifolds}

M-theory compactified
on seven-manifolds with $G_2$ holonomy also provide
$d=4$, $N=1$ vacua.  Few compact examples
are known but one may appeal to heterotic-M theory
duality in seven dimensions to make some arguments
about their structure \refs{\gtwofiber,\gtwochiral}.
For another related construction see \cvetic.

Calabi-Yau threefolds with geometric mirror partners
are believed to be $T^3$ fibrations over an $S^3$
base \syz.  Now heterotic string
theory on $T^3$ is dual to M theory on $K_3$, so
if the base is large and we stay away from
the singular fibers, we can claim that the
heterotic string on a Calabi-Yau threefold
is dual to M-theory on some $K_3$-fibered
manifold with an $S^3$ base, and hope that
the story continues when the singular $T^3$
fibers are included \refs{\gtwofiber,\gtwochiral}.
Indeed, noncompact examples which realize gauge theory
with chiral matter take the form of an ALE
space (a noncompact $K_3$) fibered over $S^3$
or over $S^3/\IZ_n$ \gtwochiral.
We will assume that sensible
compact $G_2$s exist which are $K_3$ fibrations
over $S^3/\IZ_n$.

Begin with M theory on a singular $K_3$
surface with volume $V_{K3}$.
The GUT group in such models arises from the
singularities in the $K_3$ fiber, and so one
begins with a seven-dimensional gauge
theory with dimensionful gauge coupling
$g^2 = \tl11^3$.  If we fiber this over $S_3/\IZ_p$
with volume $V_{S^3}$ then discrete Wilson lines
can break the GUT group to the
standard model at the scale $\mg = V_{S^3}^{-1/3}$.

The four-dimensional GUT coupling is
\eqn\gtwocouple{
    \ag = \frac{1}{25} = \frac{1}{4\pi V_{S_3}m_{11}^3}\ ,
}
while the four-dimensional Planck mass is:
\eqn\gtwoplanck{
    m_4^2 = \frac{V_{K_3} V_{S_3}}{2\k11^2}\ .
}

Again we can use these to fix the eleven-dimensional
Planck mass and the volume of the $K_3$ fiber:
\eqn\gtwoscales{
\eqalign{
    m_{11} &\sim \mg \cr
    \m11 & \sim 6\ \mg \cr
    V_{K3}^{-1/4} & \sim 0.1\ \mg \sim
    \frac{1}{7} \mg\ .
}}
The eleven-dimensional Planck scale and the
compactification scales are within an order of magnitude of
each other.

These compactifications look
like ``brane world'' models; the
gauge dynamics are localized on singularities
with codimension four.
One can have several singular regions in the $K_3$
fibers separate by a length of order
$V_{K_3}^{1/4} > 1/\mg, 1/\m11$.  The
singularities give rise to
$7d$ gauge dynamics and the different
gauge sectors will be ``hidden'' from each other, communicating
via 11d gravity. Furthermore, the
chiral matter also resides on singularities
which are points on $S^3$ \gtwochiral.
In the $K3$ directions, maximal supersymmetry
will be most strongly broken at the singularities
on which the gauge dynamics reside.  One can
imagine an argument similar to that in
\refs{\tomcosm,\tomcosmrev}
for the existence of inflaton candidates.
We leave this for future work.

\subsubsection{Corrections to the propagator}

For $G_2$ manifolds, the correction in \formofcorr\
is still given by \hwfoc.  Using the value
of $m_{11}$ given by \gtwoscales\ the effect is
only slightly
larger, roughly by a factor of $2$.
Again, we can ask what happens if we give up
grand unification.  Here the constraint on
$m_{11}$ is simply:
\eqn\gtwolowscale{
    m_{11}^6 = \frac{m_4^2}{50\pi V_{K_3}}\ .
}
The volume of the $S^3/\IZ_n$ base is restricted
by $\alpha \sim 1/25$.  The constraint
\eqn\kthreeconstr{
    V_{K_3} \sim \frac{(2\pi)^8}{H^4}\ ,
}
and the constraint that the
eleven-dimensional energy density be sub-Planckian,
lead to the same lower limit on $m_{11}$ to within
a factor of $2/3$.  Using the (tighter) constraint \kthreeconstr\ ,
 we find, using $H= 7 \times 10^{13} {\rm GeV}$:
\eqn\gtwonewcorr{
\eqalign{
    M &\sim m_{11} \sim 8 \times 10^{13}\ {\rm GeV}\cr
    \frac{1}{V_{K3}^{1/4}} &\sim 2 \times 10^{12}\ {\rm GeV}\cr
    \frac{H^2}{M^2} &\sim 1\ ,\ \ \ \ \ \CX \sim 0.1\ .
}}
M-theory in this limit could have an
observable effect on CMBR anisotropies,
via the corrections in \formofcorr. We emphasize
again the imprecision of our estimate of $\CX$;
it is easy to imagine gaining or losing
an order of magnitude in an explicit model.

\subsec{Type I models}

In type I models, supersymmetric
coupling unification is consistent
with weak string coupling.
As discussed in \S3, the corrections in \formofcorr\
should be computable via string perturbation
theory.  These corrections
are unobservable as long as string theory
is in a computable regime.
To see this, we will estimate the maximum size
of tree-level and one-loop contributions
to \formofcorr\ regardless of unification constraints.

A four-dimensional $N=1$ supersymmetric
model arises in type I string theory from
compactification on a six-dimensional
Calabi-Yau manifold $X$.
Tree-level corrections to $c_{1,2}$ in \effact\
are the result of compactification.  No such
terms exist in theories with sixteen supercharges.
However, higher-derivative terms
such as $R^4$ terms do exist, suppressed by
powers of $\alpha'$.  Upon compactification,
such higher-derivative terms will lead to
\eqn\treelevelcorr{
    c_{1,2} \sim
    \frac{m_{CY}^2}{\ms^2} +
    \CO \left(\frac{m_{CY}^4}{\ms^4}\right)\ ,
    \ \ \ \ \ M \sim \ms
}
in \effact, where $m_{CY} \sim V_X^{-1/6}$ is the radius
of curvature of $X$.  These terms
lead to corrections in \formofcorr\
with $M = \ms$ and $\CX$ a function
of $m_{CY}^2/\ms^2$.

The scales and couplings are constrained by:
\eqn\typeonecouplings{
\eqalign{
    \alpha &= \frac{\gs}{4 \pi \ms^6 V_6} \cr
    m_4^2 &= \frac{2 \ms^8 V_6}{\gs^2}\ .
}}
Combined, these imply:
\eqn\stringvsPlanck{
\eqalign{
    \ms^2 & = \frac{\gs m_4^2}{8\pi \alpha}\cr
    \gs & = 4\pi \alpha\left(\frac{\ms}{m_{CY}}\right)^6
}}
If the ten-dimensional
coupling is weak, then $m_{CY} \gg \ms$
and the $\alpha'$ expansion breaks down \dsscales.
If the $\alpha'$ expansion is good,
the ten-dimensional string coupling
is strong and \stringvsPlanck\ implies $\ms \geq m_4$.
In the scenario which is closest to computable,
$m_{CY} \sim \ms \sim m_4$.  The correction
terms in \formofcorr\ will appear as:
\eqn\treeratio{
    M \sim \ms \sim m_4\ ;
    \ \ \ \ \ \frac{H^2}{M^2} \sim 3 \times 10^{-8}\ ,
}
which is unobservable; a large $\CX$
in \formofcorr\ would be unnatural.
For some type I
compactifications with $m_{CY} \sim \ms$,
such as orbifolds or Gepner models,
or marginal perturbations of them,
there is hope of doing a
controllable calculation;
indeed if $m_{CY} = 2 \ms$,
$\gs \sim 10^{-2}$.
However, unless such models
deliver an extremely large value of $\CX$, which
seems unlikely, the constraints in \stringvsPlanck\
require $m_{CY}$ to be
an order of magnitude larger than
$m_s$ before $\CX H^2/\ms^2$ is observable.

We conclude that for corrections in
\formofcorr\ to be observable in a string model,
either the $2d$ $\sigma$-model coupling or
$\gs$ must be large.

\subsec{Models with TeV scale gravity}

We can take the \Hor-Witten philosophy regarding
the four-dimensional Planck scale to
a more extreme conclusion.
If we assume fewer extra dimensions,
with the standard model particles still
confined to a $3+1$-dimensional submanifold,
we may substantially
reduce the fundamental scale of quantum gravity,
as low as $m_* = 1 $ TeV, while
keeping the four-dimensional Planck scale at its known
value \add.  In particular, if there are two
extra dimensions, the compactification volume
could be as large as $(1 mm)^2$.

In this class of models the vacuum energy cannot exceed
the fundamental scale $m_*$.  Hence after the extra dimensions are
stabilized, and the effective 4D Planck scale is given by
its low energy value $m_4 \simeq 2 \times 10^{18}$~GeV,
if the vacuum energy is localized to the branes the Hubble scale
$H = \frac{m_*^2}{3\rpm}$ is incredibly small \addcosmprob,
and the mass of the inflaton must be tiny \addcosmprob,\kalin,
sixteen orders of magnitude below $m_*$. In addition the effect,
which is a correction on the order of
$H^2/m_*^2 \sim m_*^2/\rpm^2$ to the
inflaton fluctuation $\delta\phi$,
is unobservable. However, it is inconsistent to search for
inflation after such large extra dimensions are stabilized,
because if the fundamental scale
is low, inflationary dynamics after the stabilization of extra
dimensions fails to solve the age problem and does not reproduce
the spectrum of primordial fluctuations \kalin.

These problems are ameliorated if the extra dimensions play an
active dynamical role in the early universe. Specifically, if the
compactification volume was much smaller at the time of inflation
\addcos, the instantaneous Planck scale at the time of inflation
was much smaller than its later value after the stabilization,
implying that inflation at times before the extra dimensions are
stabilized can address both the age and the fluctuation problems.
The details of the pre-stabilization inflationary dynamics are
given in \addcos. The simplest realization of the scenario proposed
in \addcos\ is if the modulus parameterizing the size of the extra dimensions
itself is the inflaton.
In that case the slow roll condition can
be restated as a bound on the
ratio of the expansion rate of the dimensions
transverse to the brane (extra dimensions) to the expansion rate
of the dimensions longitudinal to the brane (macroscopic
dimensions). Representing the former by a scale factor $b$ and
the latter by $a$, it is convenient to quantify the slow roll
condition $H_b/H_a \ll 1$ (where $H_a = \dot a/a$ etc) by the
parameters $S,T$, defined by $H_b/H_a \simeq S + T (b/b_I - 1)^2 +
...~$. Here $b_I$ is the initial size of extra dimensions. Then the
slow roll conditions (i.e. the requirement to get sufficient
number of efoldings $ \ge 70$) and the scale-invariance of the
spectrum of fluctuations require $T \ll S < 0.002$ \addcos. Since
the Planck scale at the time of inflation is $m^2_{4,early} \simeq
m^{n+2}_* b^n_I$, the Hubble rate can be expressed as
\eqn\hubinfl{ H_a^2 \simeq {V \over 3 b_I^n m_*^{n+2}}} where $n$
is the number of extra dimensions and $V$ the inflationary
potential. The COBE normalization of the density contrast at $50$
efolds before the end of inflation requires \eqn\initcond{ b_I^n
m^n_* \simeq {10^3 \over S} {\sqrt{V} \over m^2_*}} and we find
after simple algebra \eqn\rairat{ \frac{H^2}{m_*^2} \simeq {S
\over 3000} {\sqrt{V} \over m_*^2} }
independently of the number of extra dimensions.
The precise value of $V$ and $S$ is clearly model-dependent;
in principle, $V$ which is supported by the branes
can be as high as $m^4_*$ and $S < 0.02$.
Thus the maximal value of the imprint of large extra
dimensions in the sky is
\eqn\maxratio{
\frac{H^2}{m_*^2} \le 6.6 \times 10^{-6} }
This is several orders of magnitude too small to be detectable.
We should stress that this formula is quite general, because it does not
depend on the number of extra dimensions nor the details of
the potential, but holds merely as a consequence of
a very basic slow roll requirement. The only assumption
which this is based on is that the radius modulus is the inflaton.
In those cases rapid asymmetric inflation
erases any short distance physics imprints on the sky
very efficiently. These conclusions might be altered
by the construction of more complex scenarios where the inflaton
is different from a radius modulus, or where the potential is
distributed throughout the bulk. However
in the case of $TeV$ gravity models, direct searches for signatures
of the new physics in colliders would be much more promising
than the surveys of the sky anyway.

\newsec{Modification of Inflationary Consistency}

A useful test of inflationary dynamics is the so-called
``consistency condition", which relates the ratio of amplitudes
of the tensor and scalar modes to the tensor tilt
(for a review of potential reconstruction and
the consistency condition, see \reconsrev).  In
standard inflation models (assuming Einstein gravity)
the spectrum of scalar fluctuations
$A_S$ determines the inflaton potential, and one can then,
in principle, use the potential reconstructed from this data
to predict the tensor spectrum $A_T$.  In practice, if one
expands $\ln(A_S)$ and $\ln(A_T)$ in a power series in
the momentum $\ln(k)$,
one can only determine the first
few coefficients in the series.
However, these are enough to provide at least a lowest-order (in the
slow-roll parameters) check
of consistency.
As we will demonstrate, if the effects of high-scale
physics are included, the usual relations
for inflation in Einstein gravity
will not be satisfied.

\subsec{Consistency in standard inflation}
We define the (unmodified by high scale physics)
scalar and tensor spectra \scaspect \tenpow
\eqn\scalspec{
 A_{S_0}(k) \equiv { H^2 \over 10 \pi m_{4}^2 H'}
 }
and
\eqn\tenspec{
A_{T_0}(k) \equiv { H \over  \sqrt{20} \pi m_{4}}.
}
Recall that the $k$ dependence is implicit in $H$.
Then
\eqn\rattenscal{
\left( A_{T_0} / A_{S_0} \right)^2 =
{2 m_{4}^2}{H'^2 \over H^2} \equiv
\epsilon_0.}
The tensor tilt is
\eqn\tentilt{
n_{T_0} \equiv {\partial (\ln{A_{T_0}^2}) \over \partial (\ln{k})}
= 2{ \partial \phi \over \partial (\ln k)} { \partial H \over \partial \phi}
= -{4 m_{4}^2 H'^2 \over  H^2}  = -2 \epsilon_0,
}
to lowest order in $\epsilon_0$.  Hence the lowest order prediction of
inflationary consistency is:
\eqn\lowconst{
n_{T_0} + 2 \left( A_{T_0} / A_{S_0} \right)^2 = 0.
}

\subsec{High-scale modifications to consistency}

When we include the effects of
high-scale physics, the observed scalar and
tensor spectra will be modified:
\eqn\spectcor{\eqalign{
&A_S = A_{S_0} (1 + \CX_S H^2 / M^2) \cr
&A_T = A_{T_0} (1 + \CX_T H^2 / M^2),}}
where $\CX_S$ and $\CX_T$ are numerical constants
in the effective action, and $M$ is the scale of
the new physics.  The ratio of the observed
tensor and scalar power spectra is:
\eqn\ratscaltencorr{
\left( A_{T} / A_{S} \right)^2 \cong
\epsilon_0 (1 + 2(\CX_T - \CX_S) H^2/M^2) \equiv \epsilon.
}
However, the tensor tilt $n_T$ is now
\eqn\tentiltcorr{
\partial (\ln{A_T^2}) / \partial (\ln{k}) = -2 \epsilon_0
\left( 1 + 2 \CX_T H^2/M^2 \right),}
and therefore
\eqn\consistcorr{
n_T + 2 \left( A_{T} / A_{S} \right)^2 = -2 \epsilon_0 \CX_S H^2/M^2 \neq 0.
}
Hence we can parameterize the predicted effect of the high-scale physics
in a way that is independent of the modification to the graviton
kinetic term.

Of course, in standard inflationary models
the consistency relation \lowconst\ is modified
at higher order in the slow roll parameters:
\eqn\highconst{
n_{T_0} + 2 {A_{T_0}^2 \over A_{S_0}^2} =
2 {A_{T_0}^2 \over A_{S_0}^2} \left( {A_{T_0}^2 \over A_{S_0}^2}
- (1 - n_{S_0}) \right) =
-2 \epsilon_0 \left( 2 \eta_0 + 3 \epsilon_0 \right),}
where $\eta$ was defined in \slowrolpar\
and $n_{S_0}$ is the unmodified scalar tilt.
As is manifest, this correction is determined by
the slow roll parameters, which in turn can be determined
via the measured scalar and tensor power and tilt.

It will therefore be possible to observe
the violation of the consistency condition
due to high scale physics if
the measurements of the scalar and tensor power
and tilts are precise enough.  This accuracy is
limited by cosmic variance, instrumental noise,
and backgrounds.  Since the tensor fluctuations
have not yet been detected, it is not known what
the backgrounds will be, and therefore how many
independent data points will ultimately be available.

Let us assume optimistically that in a region where the signal
is within a factor of three of the maximum we will be able to measure the
B-mode of the CMB polarization to an accuracy limited only by cosmic variance.
This gives
a baseline in the spherical harmonic
$l$ from, say, $l = 50$ to $l = 150$
(see {\it {e.g.}} \selzal ).
Given that the cosmic variance error in
each point is $ \sim \sqrt{l} \sim 10 $, and we have
$ \sim 100$ points, we expect
a precision of $ \sim \pm 1 \% $.  We should emphasize
again that  this estimate is close to a best case scenario.
Many other factors could stand in the
way of making cosmic variance limited measurements
of this quantity, which we must bear in mind, has not even been detected yet.

Assuming one percent precision we can continue our discussion of observability.
The violation of the consistency condition
is $2 \epsilon_0 \CX_S H^2/M^2$ \consistcorr.
If we assume $\epsilon_0 \sim 1/15$, and
$\CX_S H^2/M^2 \sim .1$, the effect will
be on the edge of observability.  Note that this
value of $\epsilon_0$ is about
the largest allowed by current experiment.
{}From the standpoint of chaotic inflation models,
it requires that the potential near the minimum
be controlled by a fairly high order monomial
$\phi^n$, which requires significant
fine tuning.  A large discrete symmetry
group at the minimum would be required
to make this technically natural.

While the corrections in \formofcorr\ lead to violations
of \highconst, they are not the only possible source of
such an effect.  In particular, they are possible in
theories with multiple scalar fields in
addition to the inflaton \refs{\multif,\corrconsist}.
Upon examining these models we find that
appreciable effects require fine tuning
of the scalar potential and the initial conditions
for the scalar fields.  In general these additional effects
are probably unobservable even with our generous estimate
of the accuracy of future experiments.

\lref\turner{M. Turner, ``Detectability of
Inflation Produced Gravitational Waves,"
{\it Phys. Rev.} {\bf D55} (1997) 435, astro-ph/9607066}

\newsec{Conclusions}

\lref\stinebring{D.R. Stinebring, M.F. Ryba, J. H. Taylor, and R.W. Romani,
``The Cosmic Gravitational Wave Background: Limits from Millisecond Pulsar Timing,"
Phys. Rev. Lett. 65 (1990) 285.}  

Let us summarize the main points we have made.
We reviewed the basics of slow roll inflation, emphasizing
in particular that the size of CMB fluctuations
is determined by inflaton fluctuations at momentum scales
$\sim H$.  The experimentally important regime is $H \ll m_4$.
It is very plausible, then, that one can express the
inflaton dynamics at scale $H$ in an effective
local field theoretic action where all unknown
short distance physics will be encapsulated
in the coefficients of irrelevant operators.
There are just two leading irrelevant operators
(see \effact ) which produce corrections to
$\delta \rho / \rho$ of size $\CX H^2 / M^2$
where $M$ is the mass scale of the short
distance physics and $\CX$ is a numerical constant
that is calculable if the short distance
physics is under calculational control.
This is one of the major results of our work.

We then turn to evaluating the size of these
corrections in various contexts.   For all renormalizable
field theories these corrections are generically
of size $H^2/m_4^2 \sim 10^{-11}$, too small to
observe (although fine tuning can make them
much larger).  Weakly coupled string theories of
conventional type display corrections of
similar size.   Regions of parameter space that
display larger corrections  clearly must involve
smaller fundamental mass scales.
Horava--Witten theory compactified with scales appropriate to
grand unification \witstrong\ has a
fundamental mass scale--the eleven dimensional Planck mass--that is
smaller,
$m_{11} \sim 5 \times 10^{16}$~GeV.   Nonetheless for this theory
$\CX H^2 / M^2 \le 10^{-7}$, still far too
small to be observed.

If we allow ourselves to give up precision grand
unification we found in section 4. that by studying
$G_2$ compactifications of M theory with large K3
fiber we could lower $m_{11}$ until it was almost
comparable to $H$, giving effects of order one
that are potentially observable.
We must stress again that these models are
not particularly attractive phenomenologically.
Precision grand unification must be abandoned,
although a reasonably large desert (up to $\sim 10^{13}$~GeV)
can be maintained.  To maintain the height of
the inflationary potential at $10^{16} ~ {\rm GeV}$ as $m_{11}$
is lowered we must abandon $m_{11}$ on the
branes as the source of  $V$ \tomcosm\ and invoke a more
recondite four dimensional mechanism that
is invariant under increases in compactification scale.
To avoid large proton decay rates we must
invoke additional discrete symmetries.

Nonetheless we believe that there are a set of viable
M theoretic models that produce potentially
observable signals.   The comments in Section 5
indicate that we will need cosmic variance limited
observations of tensor fluctuations--a very challenging,
long term experimental goal--to see such signals.
However, there may be other ways to probe these effects in the future.
Direct detection
of relic gravitational waves (by more sensitive successors to LISA, for
example)  would probe short wavelengths and so would
not be limited by cosmic variance (see {\it e.g.} \turner).   Millisecond
pulsar timing measurements would have the same advantage \stinebring . \foot{We thank E.
Witten for bringing these methods to our attention.}

Perhaps the most important lesson we have drawn
from our work is a qualitative one: the idea of probing short
distance physics using cosmological observations
looks feasible, possibly even at energies as high as
$10^{13}-10^{14}$~GeV.  The challenge now is to open the window wider.
\vskip1cm
\centerline{\bf{Acknowledgements}}
It is a pleasure to thank C. Armendariz-Picon, R. Brandenberger,
S. Carroll, S. Church, S. Dimopoulos, W. Fischler,  M. Green, B. Greene,
S. Kachru, D.E. Kaplan, A. Kosowsky, D. Langlois, A. Linde, J. McGreevy,
S. Sethi, G. Smoot,
A. Strominger, L. Susskind,  S. Thomas, E. Witten and
J. Yokoyama for valuable discussions. A.L.
would like to thank the
Aspen Center for Physics and the
High Energy Theory Group at the University of Chicago
for their hospitality at various
times during the course of this work.
This work is supported in part
by  NSF grant PHY-9870115 and by the
Stanford Institute for Theoretical Physics.
Albion Lawrence is also supported by the DOE
under contract DE-AC03-76SF00515.
Matthew Kleban is the Mellam Family Foundation Graduate Fellow.

\listrefs
\end